\documentstyle[aaspp4,12pt]{article}
\begin{document}

\title{Radial Temperature Profiles of X-Ray--Emitting
Gas Within Clusters of Galaxies}

\author{Jimmy A. Irwin and Joel N. Bregman}
\affil{Department of Astronomy, University of Michigan, \\
Ann Arbor, MI 48109-1090 \\
E-mail: jirwin@astro.lsa.umich.edu, jbregman@umich.edu}
\author{August E. Evrard}
\affil{Department of Physics, University of Michigan, \\
Ann Arbor, MI 48109-1120 \\
E-mail: evrard@umich.edu}

\begin{center}
Accepted by {\it Astrophysical Journal}
\end{center}

\begin{abstract}
Previous analyses of {\it ASCA} data of clusters of galaxies have found
conflicting results regarding the slope of the temperature profile of the hot
X-ray gas within clusters, mainly because of the large, energy-dependent point
spread function (PSF) of the {\it ASCA} mirrors. We present a summary of all
{\it ASCA}-determined cluster temperature profiles found in the literature,
and find a discrepancy in the radial temperature trend of clusters based on
which PSF-correction routine is used. This uncertainty in the cluster
temperature profile in turn can lead to large uncertainties in the amount
of dark matter in clusters.

In this study, we have used {\it ROSAT} PSPC data to obtain independent
relative temperature profiles for 26 clusters, most of which have had 
their temperature profiles determined by {\it ASCA}. Our aim is not to measure
the actual temperature values of the clusters, but to use X-ray color
profiles to search for a hardening or softening of the spectra with
radius for comparison to {\it ASCA}-derived profiles. The radial color
profiles indicate that outside of the cooling flow region, the temperature
profiles of clusters are in general constant.  Within $35\%$ of the virial
radius, we find a  temperature drop of 20\% at 10 keV and 12\% at 5 keV
can be ruled out at the 99\% confidence level. A subsample of non-cooling flow
clusters shows that the condition of isothermality applies at very small radii
too, although cooling gas complicates this determination in the cooling flow
cluster subsample. The colors predicted from the temperature profiles of a
series of hydrodynamical cluster simulations match the data very well,
although they cannot be used to discriminate among different cosmologies.
An additional result is that the color profiles show evidence for a central
peak in metallicity in low temperature clusters.
\end{abstract}

\keywords{
clusters: individual ---
intergalactic medium ---
X-rays: galaxies
}

\section{Introduction} \label{sec:intro}

The presence of hot, X-ray--emitting gas in hydrostatic equilibrium
within the gravitational potential well of a cluster of galaxies
provides a powerful tool for measuring the amount and distribution
of the total mass of the cluster (e.g., Sarazin 1988).
Once the mass of the galaxies that compose the cluster and the
mass of the hot gas is subtracted from the total mass, the amount
of dark matter contained within the cluster can be determined.
Knowledge of the amount and distribution of dark matter within clusters
is crucial for distinguishing among competing cosmological models,
that predict a high baryon fraction for a low density universe,
and vice versa (e.g., White et al.\ 1993).

The total gravitational mass contained within radius $r$ is determined
through the equation for hydrostatic equilibrium,
\begin{equation} \label{eq:hydrostatic}
M(<r) = - \frac{r k T(r)}{\mu m_H G}
\left[ \frac{d \ln \rho (r)}{d \ln r} +
\frac{d \ln T(r)}{d \ln r} \right] \, ,
\end{equation}
where the density gradient of the hot gas, $d\ln \rho(r)/d\ln r$, temperature
of the gas $T(r)$, and temperature gradient of the gas, $d\ln T(r)/d\ln r$,
need to be measured from X-ray observations. The largest uncertainty in the
mass results from $T(r)$ and $d\ln T(r)/d\ln r$. In most previous approaches,
an integrated temperature with no radial gradient was assumed.
Earlier X-ray telescopes such as {\it Ginga} and {\it EXOSAT} could
determine a global temperature, but lacked the spatial resolution needed to
determine the temperature gradient. The {\it ROSAT} PSPC possessed good
(25$^{\prime\prime}$) spatial resolution, but was not sensitive to photon
energies above 2.4 keV, making good estimates of the temperatures of hot
($5-10$ keV) clusters difficult to acquire. {\it ASCA} is sensitive to
photon energies up to 10 keV, but the large energy-dependent point spread
function (PSF) of {\it ASCA} makes it very difficult to obtain spectra within
concentric annular bins that are not significantly contaminated by photons
scattered into the bin from another bin. The average half-power diameter
of the {\it ASCA} PSF is $3^{\prime}$, and the width of the PSF increases
with increasing energy (Serlemitsos et al.\ 1995). Thus, higher energy
photons originating on-axis are more efficiently scattered to large off-axis
angular distances than lower energy photons. This has the effect of
introducing a spurious temperature increase to an isothermal profile.
For example, an isothermal profile with a temperature of 7 keV will appear
to increase to a temperature of 18 keV at an angular distance of
$20^{\prime}$ (Takahashi et al.\ 1995) due to the widening of the PSF at
higher energies. Still, in principle it is possible to account for the PSF
and derive accurate temperature profiles for clusters.

Radial temperature profiles for many clusters observed with {\it ASCA} have
been published in the literature that are reported to have been corrected
for the energy-dependent PSF (Markevitch et al.\ 1998, hereafter M98; 
Markevitch 1996; Ikebe 1995)
However, in some cases where the same data
have been analyzed by different authors, significantly different results
have been found. It is important to determine which PSF-correction
method is correct, since a nearly complete flux-limited sample of nearby
clusters has already been observed with {\it ASCA}, and it is unlikely that this
will be repeated with {\it AXAF} in the near future. Even though {\it AXAF}
has the required spatial resolution and wide energy
bandpass to determine the radial temperature gradient within clusters, the
rather small field of view of {\it AXAF} is insufficient to image
nearby clusters out to large angular distances ($\sim20^{\prime}$) without
resorting to the time-consuming process of observing the cluster with
several different pointings.

In this paper, we summarize all {\it ASCA}-derived cluster
temperature profiles found in the literature, and use archival {\it ROSAT}
PSPC data to attempt to determine whether temperature gradients exist
in the hot gas contained within clusters of galaxies.
Rather than perform detailed spectral fitting to the
{\it ROSAT} PSPC data, we calculate three X-ray ``colors" (ratios of
counts in bands covering the {\it ROSAT} bandpass) in an effort to detect
a hardening or softening of the spectra as a function of radius. This
method has the advantage of being non-parametric, and is suitable for
detecting changes in temperature, as opposed to the actual temperature values.
A technique similar to this has been used to search for cooling gas
and excess absorption in the centers of cooling flow clusters observed by
{\it ROSAT} by Allen \& Fabian (1997). These authors find that their color
profiles suggest isothermality outside of the cooling flow region, although
their profiles typically extend out to only $5^{\prime}$. Our goal is to
extend this result out to $\sim 15^{\prime}$.
In addition, this type of colors technique has also been used recently by
Ettori, Fabian, \& White (1998) to search for temperature and absorption
variations in the Perseus cluster.

\section{Previous Published Results With {\it ASCA}} \label{sec:asca_temps}

All clusters we could find in the literature observed with {\it ASCA}
with regular, symmetric X-ray
emission for which PSF-corrected temperature profiles have been determined
to date, and for which a description of the PSF-correction method has
been outlined are listed in Table~\ref{tab:asca}. We have excluded temperature
profiles of clusters with asymmetric X-ray emission, for which a radial
profile is not useful, and also clusters for which the PSF was not accounted
for in the analysis (except for a few low temperature clusters and A2218;
see below). For each cluster the trend in temperature with
{\small
\begin{table}[thbp]
\vspace{-0.3truein}
\caption[Published {\it ASCA}-derived Cluster Temperature Profiles]{}
\label{tab:asca}
\begin{center}
\begin{tabular}{ccccc}
\multicolumn{5}{c}{\sc Published {\it ASCA}-derived Cluster
Temperature Profiles} \cr
\tableline \tableline
Cluster & Radial Trend & \% Prob.\tablenotemark{a} & Method & Ref. \\
\tableline
A85 & Decreasing & $>99.99$ & 1 & M98 \\
A119 & Decreasing & 73.6 & 1 & M98 \\
A399 & Constant & 2.6 & 3 & Fujita et al.\ 1996 \\
A399 & Decreasing & 86.8 & 1 & M98 \\
A401 & Constant & 27.5 & 3 & Fujita et al.\ 1996 \\
A401 & Decreasing & 99.98 & 1 & M98 \\
A478 & Decreasing & 94.5 & 1 & M98 \\
A644 & Decreasing & 87.1 & 1 & Bauer \& Sarazin 1998 \\
A665 & Decreasing & 99.66 & 1 & M98 \\
A780 & Decreasing & 52.0 & 1 & Ikebe et al.\ 1997  \\
A780 & Decreasing & 80.5 & 1 & M98 \\
A1246 & Constant & \ldots & 2 & Yamasaki et al.\ 1997 \\
A1650 & Decreasing & 30.1 & 1 & M98 \\
A1651 & Decreasing & 79.5 & 1 & M98 \\
A1795 & Constant & 14.8\tablenotemark{b} & 2 & Ohashi et al.\ 1997 \\
A1795 & Decreasing & 49.2 & 1 & M98 \\
A2029 & Decreasing & 50.1 & 1 & Sarazin et al.\ 1997 \\
A2065 & Decreasing & 99.87 & 1 & M98 \\
A2142 & Decreasing & 74.9 & 1 & M98 \\
A2163 & Decreasing & $>99.99$ & 1 & M98 \\
A2218 & Decreasing & $>99.9$ & 4 & Cannon et al.\ 1998 \\
A2256 & Decreasing & $>99.99$ & 1 & Markevitch 1996 \\
A2319 & Decreasing & 87.9 & 1 & Markevitch 1996 \\
A2657 & Decreasing & 97.6 & 1 & M98 \\
A3112 & Decreasing & 98.1 & 1 & M98 \\
A3266 & Decreasing & 99.3 & 1 & M98 \\
A3391 & Constant & 1.6 & 1 & M98 \\
A3558 & Decreasing & 91.7 & 1 & Markevitch \& Vikhlinin 1997 \\
A3571 & Decreasing & 97.2 & 1 & M98 \\
A4059 & Decreasing & 99.75 & 1 & M98 \\
MKW3S & Constant & \ldots & 2 & Kikuchi et al.\ 1996 \\
MKW3S & Decreasing & 99.25 & 1 & M98 \\
Triangulum Australis & Decreasing & 97.5 & 1 & M98 \\
Ophiuchus & Constant & \ldots & 2 & Matsuzawa et al.\ 1996 \\
Centaurus & Increasing & 47.9 & 2 & Ikebe 1995 \\
AWM7 & Constant & 66.0 & 2 & Ezawa et al.\ 1997 \\
AWM7 & Constant & 34.1 & 1 & Markevitch \& Vikhlinin 1997 \\
2A0335+096 & Constant & 17.4 & 2 & Ohashi et al.\ 1997 \\
\tableline
\end{tabular}
\end{center}
%\tablenotetext{a}{\% Prob.\ is the percentage probability that the derived
$^a$\% Prob.\ is the percentage probability that the derived
temperature profile is inconsistent with a constant value ($dT(r)/dr=0$),
as derived from a $\chi^2$ test. \\
%\tablenotetext{b}{After exclusion of the $5^{\prime}-9^{\prime}$ annulus
$^b$After exclusion of the $5^{\prime}-9^{\prime}$ annulus
which may be contaminated by a hard point source
\end{table}
}
increasing radius is designated as increasing, constant, or
decreasing. To determine the significance of the radial trends we fitted
constant temperature profiles to the published data using the
$\chi^2$ test. The percent probability that the profile is inconsistent with
a constant temperature profile is shown in Table~\ref{tab:asca} (denoted
by \% Prob.), for those clusters for which actual temperature values were given
in the original paper.

Various methods have been developed to account for the energy-dependent PSF
of {\it ASCA}. Method 1 is described in Markevitch et al.\ 1996, Method 2
is described in Ikebe (1995), Method 3 is described in Fujita et al.\ (1996).
Method 4 (Cannon, Ponman, \& Hobbs 1998) did not correct for the PSF,
but is mentioned because it still finds a negative temperature gradient
for A2218. The method used is shown in Table~\ref{tab:asca}.

In general those clusters analyzed using Method 1
show a decreasing temperature profile (up to a factor of two), except
for A3391 and AWM7, leading to the claim of a ``universal" temperature
profile for non-merging clusters (M98). Of the 28 clusters analyzed
in this manner, 22 (14) were inconsistent with a constant temperature profile
at the 70\% (90\%) probability level.
On the other hand, all but one of the clusters analyzed using Method 2
show constant profiles, and the lone exception
(Centaurus) is consistent with a constant value at around the 50\%
probability level.

Particularly interesting are the clusters that have been analyzed using
more than one technique. Two such clusters are the non-cooling flow cluster
pair A399 and A401. When analyzed with Method 1
(M98), the temperature was found to decrease radially by
a factor of 1.5 and 1.7 for A399 and A401, respectively. Such a decline,
if real, would have a substantial effect on the total mass estimate of
the cluster. These two clusters were also analyzed with Method 3 by
Fujita et al.\ (1996), who found a flat temperature profile for both clusters.
For their spectral fitting, Fujita et al.\ (1996) used ancillary
response files that took the energy-dependent PSF into account,
but which assumed a constant spectrum for the entire field when calculating
the amount of contamination from other regions of the detector. Although this
would have no effect on an isothermal profile, it will tend to flatten
any existing temperature gradient. It is unlikely, though, that this effect
can make a temperature profile that decreases by a factor of 1.5 to
appear flat (Fujita 1998). A similar discrepancy also exists for MKW3S.
When analyzed with Method 1 (M98), a decrease from 4 keV
to 2.8 keV with small errors was found, whereas a constant profile was found
using Method 2 (Kikuchi et al.\ 1996). A1795 was found
to have a flat profile using Method 2 (Ohashi et al.\ 1997)
and a decreasing (although not significant) profile using
Method 1 (M98). The other cluster that has been analyzed
by the two techniques shows agreement between the two techniques; AWM7 has
a constant profile.

Temperature profiles of several low temperature ($kT < 4$ keV) clusters
that have not been corrected for the PSF of {\it ASCA} have also been presented
in the literature. At these low temperatures the spurious temperature
increase caused by the PSF is minimal (Takahashi et al.\ 1995). A1060
(Tamura et al.\ 1996), A262 and MKW4S (Fukazawa et al.\ 1996), and
A496 (Hatsukade \& Ishizaka 1996) all show a constant temperature profile
outside of the cooling flow region.

Given the apparent discrepancy between various techniques, it is important
to determine which technique gives a more accurate result. {\it ROSAT}
PSPC data might shed some light on the matter. Although not sensitive
to photon energies above 2.4 keV, large (factor of two) gradients in the
temperature profile should be detectable with {\it ROSAT}.

\section{Sample Selection} \label{sec:sample}

For our sample, we chose selection criteria similar to those used by
M98, who analyzed 30 clusters observed with
{\it ASCA}. This was done to ensure that a direct comparison could be made
between most of the clusters in our {\it ROSAT} PSPC sample and clusters
analyzed with {\it ASCA}. We chose Abell
clusters that fell in the redshift interval of $0.04\le z \le 0.09$,
and had {\it ROSAT} fluxes greater than $2 \times 10^{-11}$ ergs s$^{-1}$
cm$^{-2}$ (Ebeling et al.\ 1996), plus the clusters Cygnus A, MKW3S and
Triangulum Australis. At this redshift range, most of the X-ray emission
from the clusters fit comfortably within the rib support
structure of the {\it ROSAT} PSPC at an angular distance of 18$^{\prime}$.
Of the 36 clusters that satisfied our selection criteria, 32 were
the targets of pointed observations with the {\it ROSAT} PSPC. Of these
32 clusters, we discarded 6 clusters that exhibited very asymmetric X-ray
emission, for which a symmetrical radial profile has no real meaning. The
26 clusters in our sample are listed in Table~\ref{tab:sample}, along with
a few relevant properties of each cluster. The temperatures given were taken
from the literature and were mainly determined with {\it ASCA} data,
although a few were measured with {\it Einstein}, {\it ROSAT},
{\it Ginga}, or {\it EXOSAT}.
\begin{table}[tbhp]
\caption[Sample of {\it ROSAT} Observed Clusters]{}
\begin{center}
\begin{tabular}{cccc}
\multicolumn{4}{c}{\sc Sample of {\it ROSAT} Observed Clusters} \cr
\tableline \tableline
Cluster & Redshift & Exposure (s) & Temperature (keV) \\
\tableline
A85 & 0.052 & 13361 & 6.9 \\
A119 & 0.044 & 12900 & 5.6 \\
A133 & 0.060 & 15055 & 4.0 \\
A401 & 0.074 & 11772 & 8.0 \\
A478 & 0.088 & 21239 & 8.4 \\
A644 & 0.071 & 8817 & 7.9 \\
A780 & 0.057 & 16407 & 4.3 \\
A1651 & 0.085 & 7153 & 6.1 \\
A1795 (high gain) & 0.062 & 24874 & 7.8\\
A1795 (low gain) & \ldots & 33644 & \ldots \\
A2029 & 0.077 & 12602 & 9.1 \\
A2142 (PSPC-B) & 0.089 & 14608 & 9.7 \\
A2142 (PSPC-C) & \ldots & 7120 & \ldots \\
A2256 & 0.058 & 16104 & 6.6 \\
A2319 & 0.056 & 4003 & 8.8 \\
A2589 & 0.042 & 6510 & 3.7 \\
A2597 & 0.085 & 6551 & 4.4 \\
A2657 & 0.040 & 17110 & 3.7 \\
A3112 & 0.070 & 6774 & 5.3 \\
A3158 & 0.059 & 2834 & 5.5 \\
A3266 & 0.055 & 19536 & 8.0 \\
A3391 & 0.054 & 6228 & 5.4 \\
A3532 & 0.056 & 6761 & 4.4 \\
A3562 & 0.050 & 17055 & 3.8 \\
A3571 & 0.040 & 4950 & 6.9 \\
A4059 & 0.048 & 5165 & 4.4 \\
MKW3S & 0.045 & 7801 & 3.7 \\
Triangulum Australis & 0.051 & 6517 & 9.6 \\
\tableline
\end{tabular}
\end{center}
\label{tab:sample}
\end{table}
Using this nearly complete sample, our aim
was to select a sample that was large enough to search for trends in the
hardness ratios of the X-ray emission in clusters, even if the
statistics for a given cluster are not good enough to determine a trend
on an individual basis.

\section{Data Reduction} \label{sec:data}
For each cluster, the data were processed in the following manner. First,
we used the latest version of {\sc pcpicor} within {\sc FTOOLS 4.0} to correct
for temporal and spatial gain variations across the image plane of the
PSPC that were incorrectly handled by SASS as part of the conversion
from detected pulse height to pulse invariant channel (Snowden et al.\ 1995).
The detector coordinates and arrival time of each event are used to
correct the data for this effect.
Next, the data were corrected for particle and solar X-ray background,
exposure, and vignetting as described by Snowden (1995). Periods
of high background due to charged particles were removed by filtering
the data such that all time intervals with a Master Veto Rate above 170
counts s$^{-1}$ were excluded. Since some of the clusters lie in directions
of high Galactic hydrogen column densities, we excluded data in Snowden
(1995) bands R1, R2, and R3 (approximately 0.11--0.51 keV) since these
bands often have very low X-ray count rates due to absorption. We created
three energy bands with the remaining four energy bands created by the
Snowden (1995) routines: R4 + R5 (approximately
0.52--0.90 keV), R6 (approximately 0.91--1.31 keV), and R7
(approximately 1.32--2.01 keV). Nearly all the data were taken in the
post-gain change mode (1991 October 11) with the PSPC-B, although a few clusters
were observed before the gain change, and two with the PSPC-C.

For each of the three energy bands, counts were extracted from 3--4 concentric
annuli centered on the centroid of the X-ray emission. The outer radius of
the central bin was usually chosen to be $1\farcm5-2^{\prime}$ to encompass
all or nearly all emission from a cooling flow that might be present.
When a cooling flow was not present, the innermost bin often was made
larger. In many cases, the extraction regions were chosen to match those
of previous studies that analyzed {\it ASCA} data.
Unrelated X-ray sources in the
field of view were removed from the data. For each band, background counts
were obtained from an annulus $30^{\prime}-40^{\prime}$ in extent, and
subtracted from the source counts. With a few exceptions, the annular rings
covered the full azimuthal range. The exceptions are A85, A2256, and A2319,
which show some asymmetry possibly due to a merger. The asymmetric
portion of their X-ray emission was not analyzed.

From these three bands, we constructed three ratios or ``colors" defined as
\begin{equation} \label{eq:c76}
{\rm C76} =
\frac{\rm counts~in~1.32-2.01~keV~band}{\rm counts~in~0.91-1.31~keV~band}
\, ,
\end{equation}
\begin{equation} \label{eq:c745}
{\rm C745} =
\frac{\rm counts~in~1.32-2.01~keV~band}{\rm counts~in~0.52-0.90~keV~band}
\, ,
\end{equation}
and
\begin{equation} \label{eq:c645}
{\rm C645} =
\frac{\rm counts~in~0.91-1.31~keV~band}{\rm counts~in~0.52-0.90~keV~band}
\, .
\end{equation}
A1795 was observed in both high gain and low gain modes, so the data
from each gain mode have been analyzed separately. Similarly, A2142 was
observed with both PSPC-B and PSPC-C, so the data from each instrument were
analyzed separately.

\section{Color Models} \label{sec:color_models}

As noted in \S~\ref{sec:intro}, the limited ($<2.4$ keV) bandpass of the
{\it ROSAT} PSPC does not allow tight constraints to be placed on the
temperature of hot clusters. There are conflicting views concerning the
reliability of the {\it ROSAT} PSPC to determine accurate global values
for hot clusters because of calibration uncertainties of the instrument.
Markevitch \& Vikhlinin (1997) found
that {\it ROSAT} gave global temperatures for the clusters A3558 and AWM7
that were 1.2--1.75 times lower than the temperatures found by
{\it ASCA} and other high-energy instruments. On the other hand, Henry (1998)
has updated Table~1 of Henry \& Briel (1996) and found that
the {\it ROSAT}-derived temperatures of seven of nine clusters agreed
with the temperatures found by {\it ASCA} or {\it EXOSAT} within the 1$\sigma$
errors. In six of the nine cases, the
{\it ROSAT} temperature was less than the temperature derived by other
instruments. We choose to avoid the issue of whether the PSPC can accurately
determine the temperature of hot clusters, and employ the use of
X-ray colors rather than detailed spectral fitting to search for
temperature gradients. We are not concerned with the actual value of the
temperature, just changes in the temperature, which should not be affected by
calibration uncertainties, unless these uncertainties change as a
function of position on the detector. 
\begin{figure}[htbp]
\vskip6.30truein
\hskip0.3truein
\includegraphics{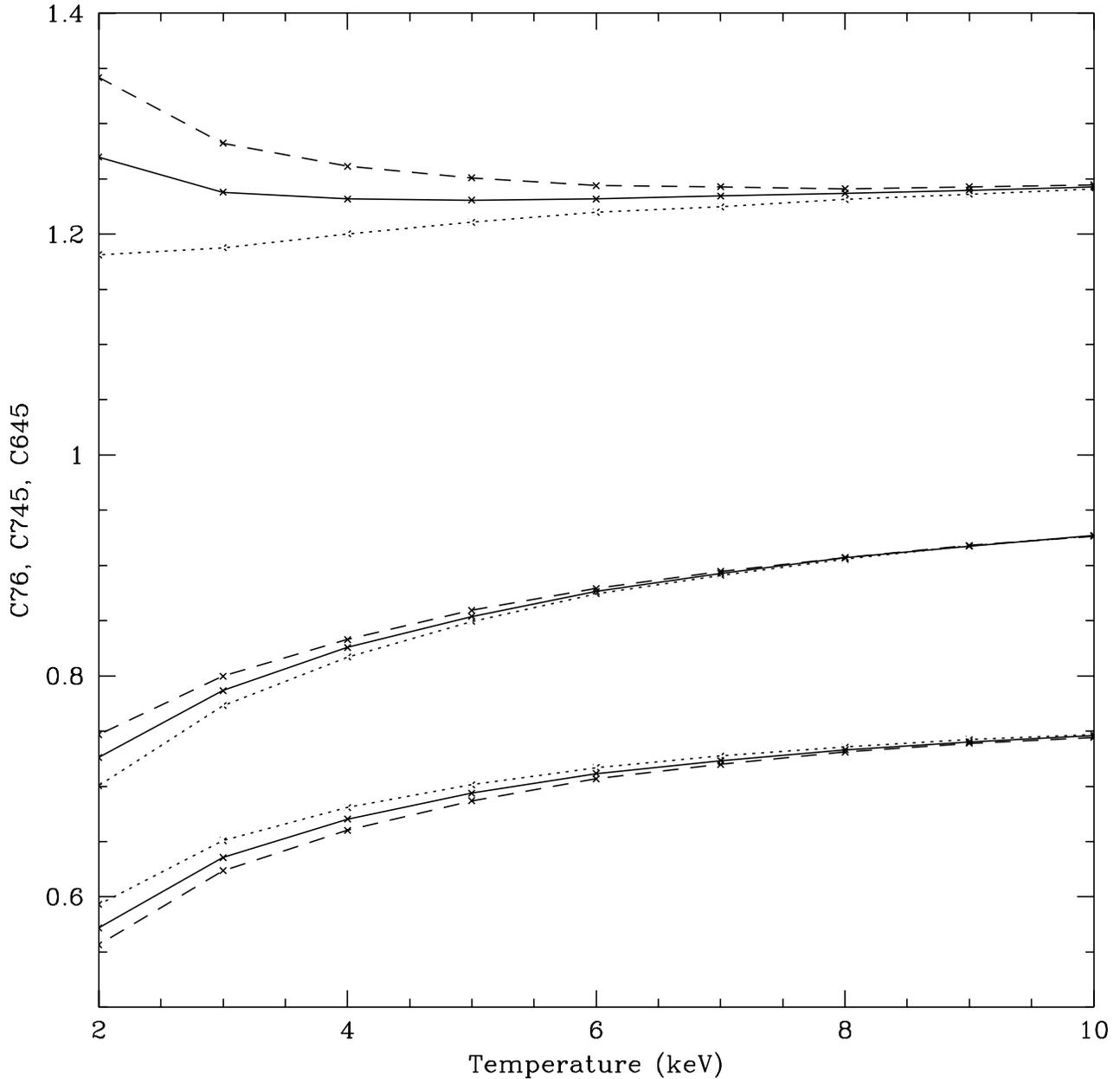}
\caption[Model Color Profiles - Temperature]{
Model color profiles derived from the PSPC-B (low gain) response for a
variety of Raymond-Smith spectral models. The top lines are C645, the
middle lines are C745, and the bottom lines are C76. For each color, dashed
lines represent models with 60\% metallicity, solid lines represent 40\%
metallicity, and dotted lines represent 20\% metallicity. Metallicity
has a small effect on C645 at low temperatures, but has a minimal effect on
C76 and C745 Note that the effect of metallicity on C745 and C76 is
reversed. These models illustrate that C76 and C745 are sensitive
to changes in temperature, whereas C645 is insensitive to changes in
temperatures above 5 keV or near 40\% metallicity.
\label{fig:model_temps}}
\end{figure}

First, we address the issue as to whether a change in temperature with
radius in the 2--10 keV range would be detectable with {\it ROSAT}.
To do this, we have calculated the colors expected from a variety of
Raymond-Smith (1977) models, once the spectral models have been folded
through the spectral response of the PSPC using XSPEC. For these models, we
have assumed metallicities of 20\%, 40\%, and 60\% and an absorbing hydrogen
column density of $5 \times 10^{20}$ cm$^{-2}$, although variations
in absorption will be discussed below. The predicted colors C76, C745, and C645
are shown for temperatures ranging from 2 keV to 10 keV in
Figure~\ref{fig:model_temps}. These plots show the color values for low gain
data. Model colors for PSPC-B high gain and PSPC-C data differed by several
percent from colors derived from PSPC-B low gain data, but were qualitatively
very similar as far as their dependence on temperature and metallicity.
The colors C76 and C745 are sensitive to changes
in temperature. For C76 the percentage decrease in going from 10 keV
to 5 keV, from 6 keV to 3 keV, and from 4 keV to 2 keV is 7.0\%, 10.7\%,
and 14.7\%, respectively, for models
with 40\% metallicity. For C745 the percentage decreases were 7.9\%, 10.2\%,
and 12.1\%,
respectively, for models with 40\% metallicity. The effect of abundance
variation on C76 and C745 is negligible above 6 keV and minimal even at 3 keV.
Therefore, these two colors appear to be a good indicator of large (factor
of 2) changes in the temperature. On the other hand, the color C645 is
insensitive to changes in temperature. At 40\% metallicity, C645 is
completely independent of temperature above 3 keV, and is only minimally
affected
by temperature above 6 keV at 20\% and 60\% metallicities. The change in
C645 in going from 6 keV to 3 keV is only 2.6\% and 3.1\% for 20\% and 60\%
metallicities, respectively. Since C645 is not sensitive to temperature
(above 3 keV),
this color is a useful test of our contaminant removal and background
subtraction procedures, as well as an indicator of absorption
gradients, if present.

\begin{figure}[htbp]
\vskip6.30truein
\hskip0.3truein
\includegraphics{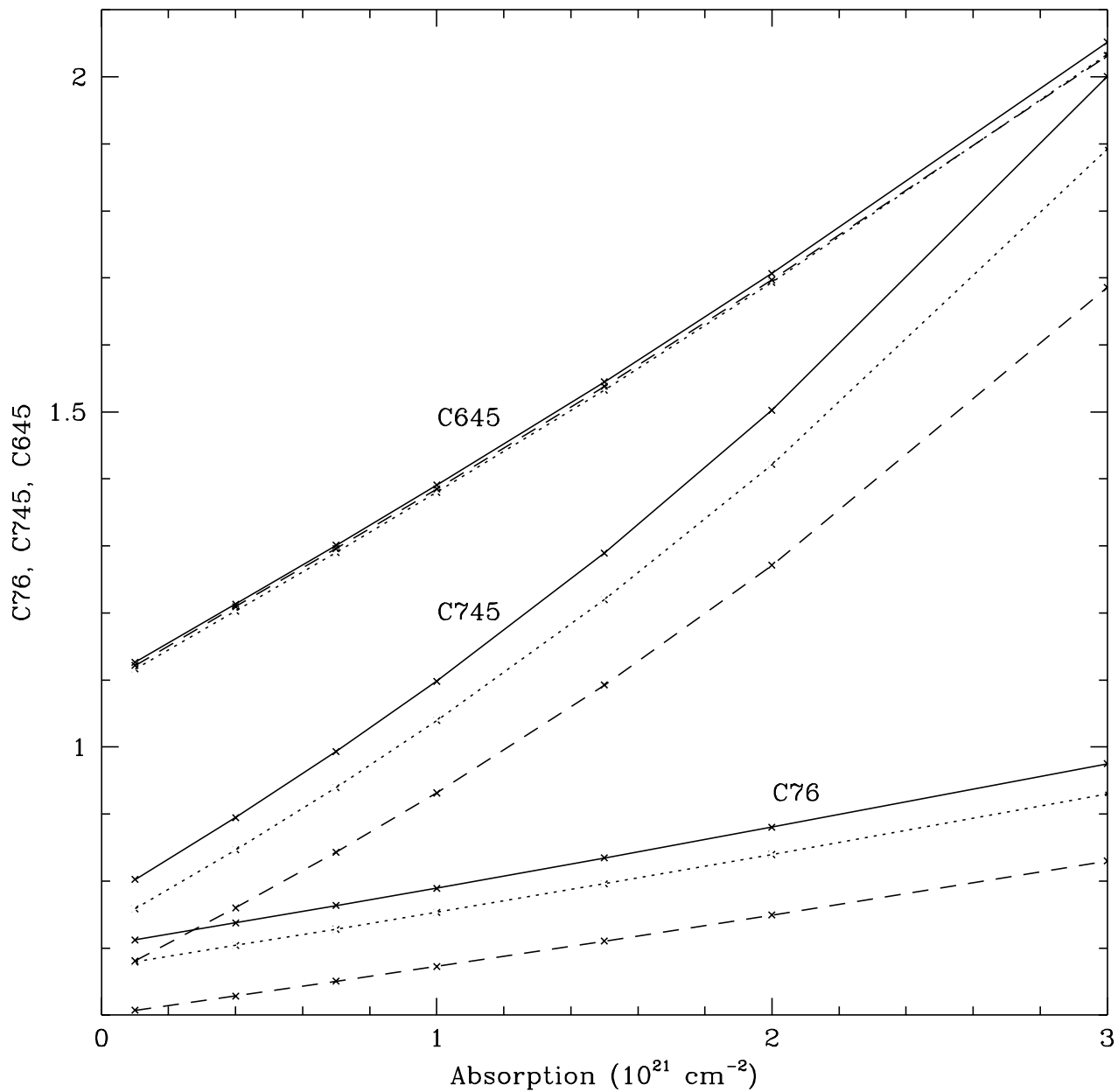}
\caption[Model Color Profiles - Absorption]{
Model color profiles as a function of absorbing column density. The top lines
are C645, the middle lines are C745, and the bottom lines are C76. The
metallicity of the models is 40\%. For each
color, the solid line represents the 10 keV track, the dotted line
represents the 6 keV track, and the dashed line represents the 3 keV track.
The relative ratios of colors at different temperatures are independent
of column density.
\label{fig:absorption}}
\end{figure}

Figure~\ref{fig:absorption} shows the effect a change in absorption has
on the colors at three different temperatures. For all three colors,
the percentage change between two temperature tracks is independent of the
absorbing column density. Thus, a change in temperature from 10 keV to 5 keV
(for example) will lead to the same percentage decrease in C76 for a cluster
with an absorbing column density of $10^{20}$ cm$^{-2}$ as a cluster with
an absorbing column density of $3 \times 10^{21}$ cm$^{-2}$. This will be
important when combining the color profiles from each individual cluster
into one composite profile (see \S~\ref{sec:discussion} below).
\begin{figure}[htbp]
\vskip6.30truein
\hskip0.3truein
\includegraphics{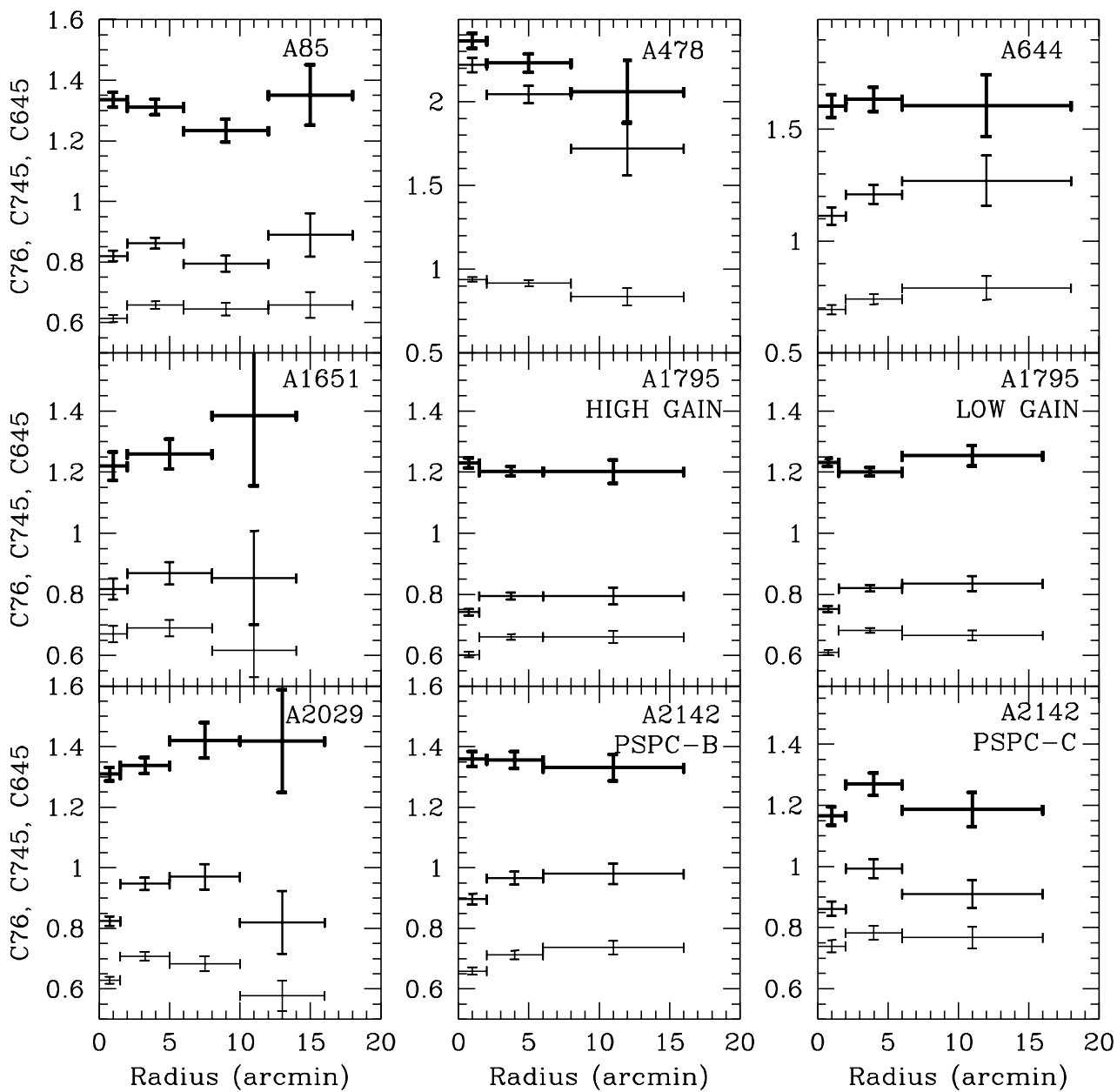}
\caption[Hot Cooling Flow Clusters]{
Radial color profiles for clusters with $kT>5$ keV and
$\dot M > 50~M_{\odot}$ yr$^{-1}$. The top profile
in each panel is C645, the middle is C745, and the bottom is C76.
\label{fig:highT-cf}}
\end{figure}
\begin{figure}[htbp]
\vskip6.30truein
\hskip0.3truein
\includegraphics{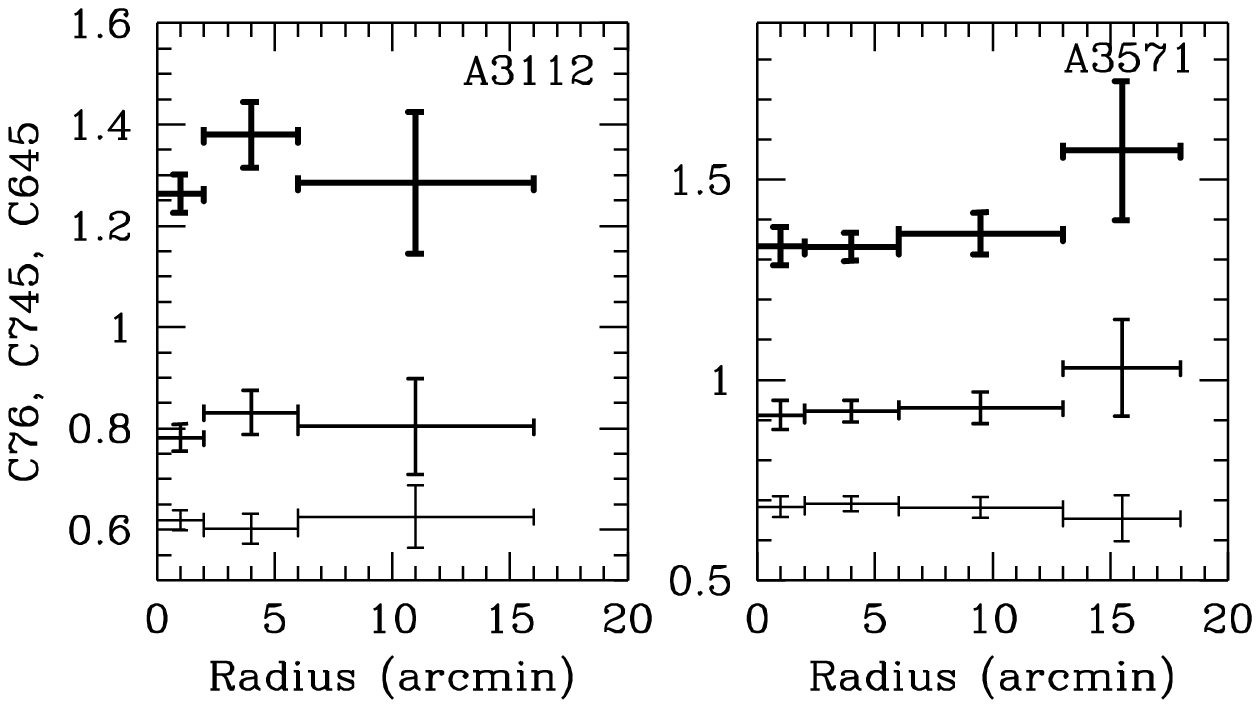}
Fig.~3 - continued
\end{figure}
\begin{figure}[htbp]
\vskip6.30truein
\hskip0.3truein
\includegraphics{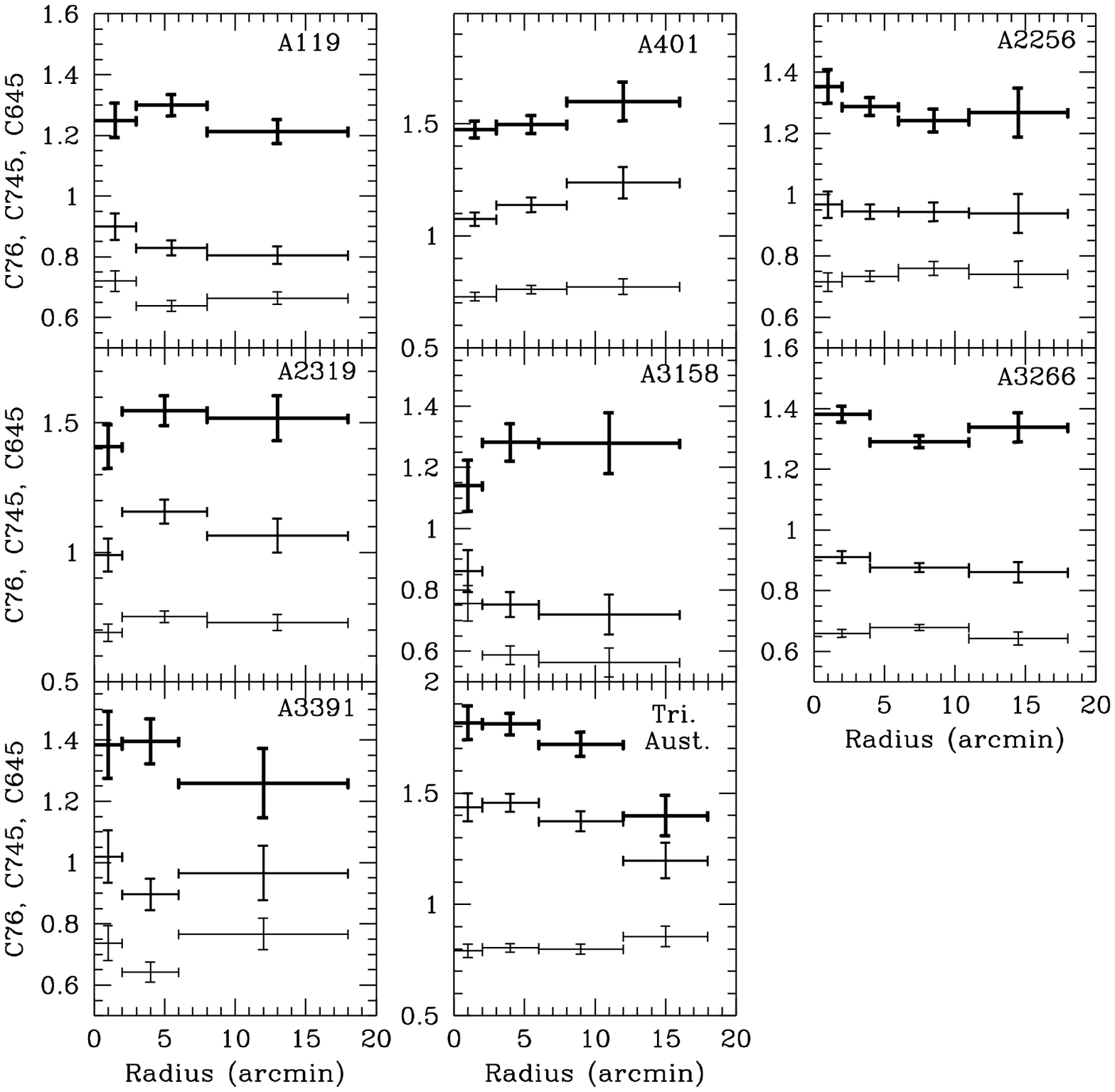}
\caption[Hot Non-Cooling Flow Clusters]{
Radial color profiles for clusters with $kT>5$ keV and
$\dot M < 50~M_{\odot}$ yr$^{-1}$. The notation is the same as that for
Figure~\protect\ref{fig:highT-cf}.
\label{fig:highT-nocf}}
\end{figure}
\begin{figure}[htbp]
\vskip6.30truein
\hskip0.3truein
\includegraphics{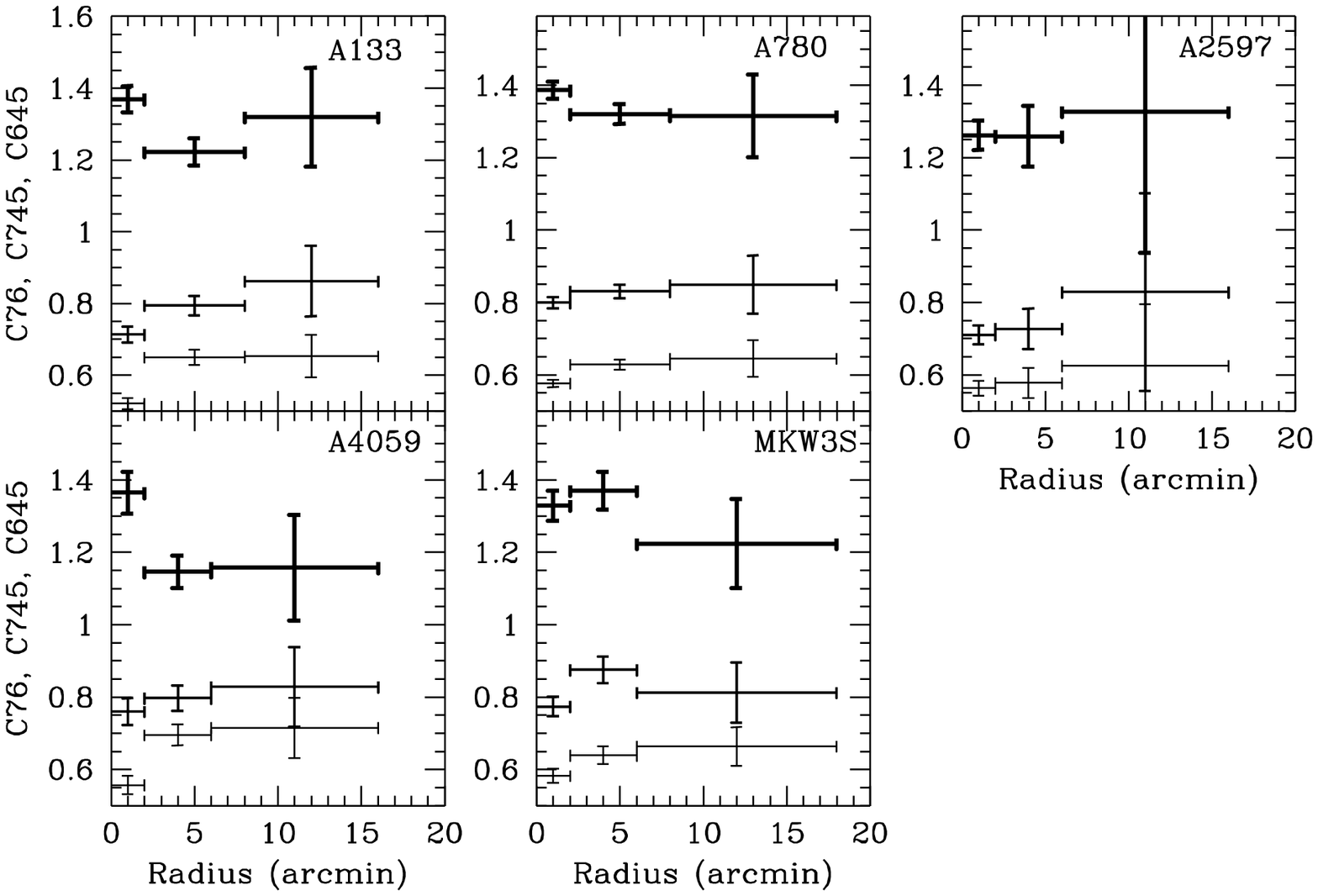}
\caption[Cool Cooling Flow Clusters]{
Radial color profiles for clusters with $kT<5$ keV and
$\dot M > 50~M_{\odot}$ yr$^{-1}$. The notation is the same as that for
Figure~\protect\ref{fig:highT-cf}.
\label{fig:lowT-cf}}
\end{figure}
\begin{figure}[htbp]
\vskip6.30truein
\hskip0.3truein
\includegraphics{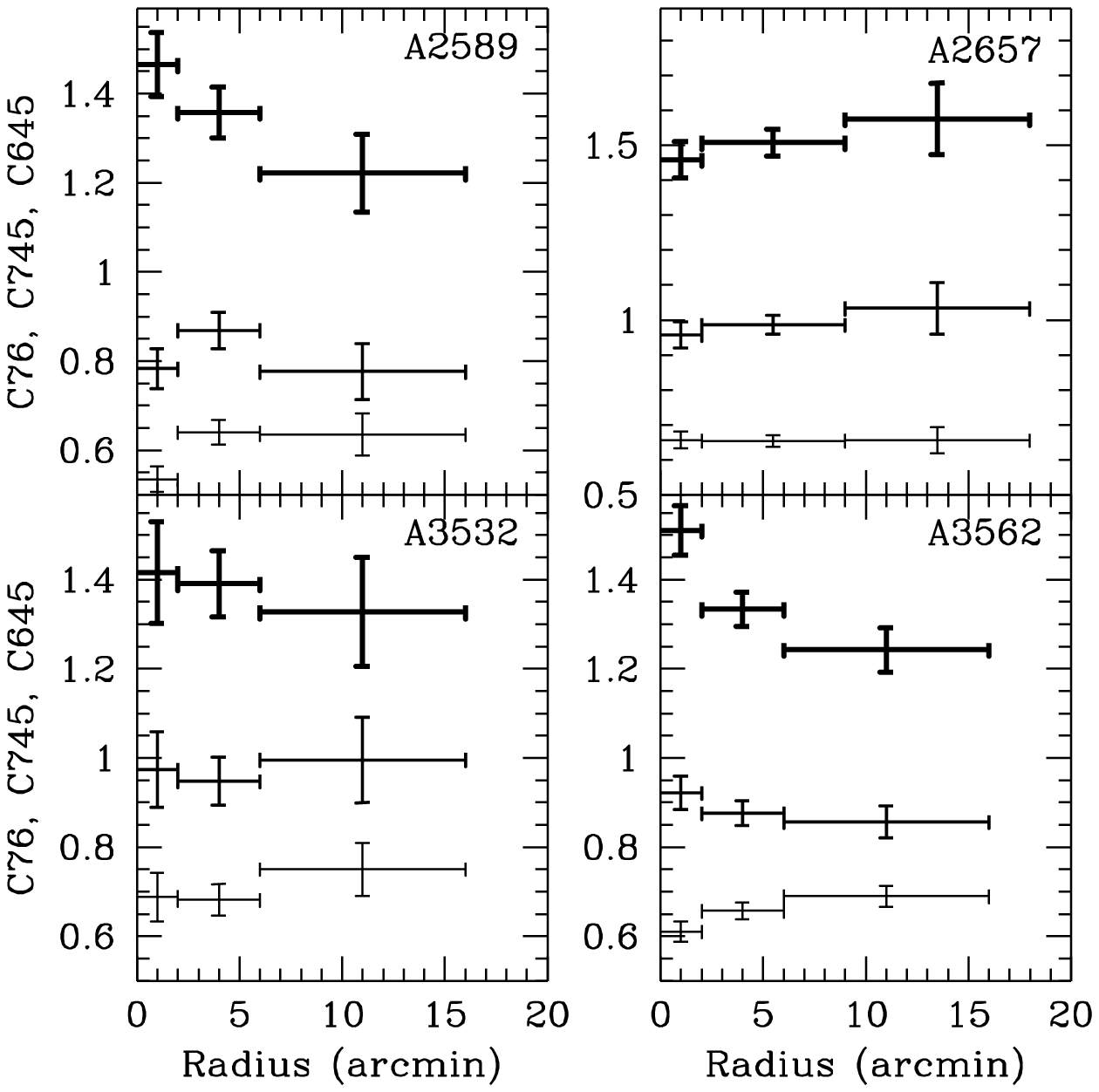}
\caption[Cool Non-Cooling Flow Clusters]{
Radial color profiles for clusters with $kT<5$ keV and
$\dot M < 50~M_{\odot}$ yr$^{-1}$. The notation is the same as that for
Figure~\protect\ref{fig:highT-cf}.
\label{fig:lowT-nocf}}
\end{figure}

Variations in the Galactic column densities across the field of view
of the PSPC should not have a significant effect on the colors.
Arabadjis \& Bregman (1998) found that for column densities below
$10^{21}$ cm$^{-2}$, the column density varies by no more than
$\sim$5\% on angular scales of $\sim$30$^{\prime}$. At
$5 \times 10^{21}$ cm$^{-2}$, a 10\% change in the column density led
to changes in C76, C745, and C645 of only 0.6\%, 1.8\%, and 1.2\%,
respectively.

\section{Observed Radial Color Profiles} \label{sec:color_profiles}

The color profiles for the 26 clusters in our sample are shown in
Figure~\ref{fig:highT-cf}--\ref{fig:lowT-nocf}. The sample has been broken
into four groups according to temperature and cooling rate:
$kT>5$ keV, $\dot M > 50~M_{\odot}$ yr$^{-1}$
(Figure~\ref{fig:highT-cf}), $kT>5$ keV, $\dot M < 50~M_{\odot}$ yr$^{-1}$
(Figure~\ref{fig:highT-nocf}), $kT<5$ keV, $\dot M > 50~M_{\odot}$ yr$^{-1}$
(Figure~\ref{fig:lowT-cf}), and $kT<5$ keV, $\dot M < 50~M_{\odot}$ yr$^{-1}$
(Figure~\ref{fig:lowT-nocf}). A2589 had no published cooling rate, so it
has been placed in the no cooling flow category (for display only).
Below we discuss each cluster briefly. The cooling rates given are from
Peres et al.\ (1998) unless otherwise noted.

\subsection{$kT>5$ keV, $\dot M > 50~M_{\odot}$ yr$^{-1}$}
\label{ssec:highT-cf}

{\it A85}: A clump in the X-ray emission to the south of the cluster
that may or may not be physically associated with the cluster
(Lima Neto et al.\ 1997)
was excluded from the analysis here. The resulting C76 and C745 profiles are
constant within the errors outside the cooling flow region
($\dot M = 198~M_{\odot}$ yr$^{-1}$) as is the C645 profile, indicating
isothermality. This is consistent with the results of Pislar et al.\ (1997)
and Kneer et al.\ (1995) who performed detailed spectral modeling of the same
data. Conversely, M98 found a decrease in the temperature of a factor of two
for this cluster with {\it ASCA} data.

{\it A478}: This cluster is the most heavily absorbed in our sample. All
three color profiles show a considerable decreasing trend, indicative of
a radial trend in the absorbing column density. This is
consistent with the results of Allen et al.\ (1993), who analyzed the same
{\it ROSAT} data and found the absorbing column density to decrease by
a factor of 1.5 from the center to the outer part of the cluster. Such
an absorption gradient apparently masks the effect of a cooling flow
($\dot M = 616~M_{\odot}$ yr$^{-1}$) on the color profiles.

{\it A644}: C76 and C745 are constant outside the cooling flow region
($\dot M = 189~M_{\odot}$ yr$^{-1}$), as is C645.
Bauer \& Sarazin (1999) found the temperature to
decrease from 10 keV in the center to 5.1 keV with {\it ASCA} data.
 
{\it A1651}: All three color profiles are constant within the (large) errors,
indicating no temperature, metallicity, or absorption gradients, although
this cluster possesses a moderate cooling flow 
($\dot M = 138~M_{\odot}$ yr$^{-1}$).

{\it A1795}: This cluster was observed in both high gain and low gain
modes, so the data were analyzed separately. In both data sets, C76 and
C745 show a drop in the center, indicative of a cooling
flow ($\dot M = 381~M_{\odot}$ yr$^{-1}$). The outer two bins
of C76 and C745 are consistent with small errors. The C645 profile is
also constant with small errors. Given the excellent statistics of both
observations, there is little doubt that this cluster is isothermal
outside of the cooling flow region. Briel \& Henry (1996) found a nearly
constant temperature profile outside the cooling radius with the same
{\it ROSAT} data.

{\it A2029}: This cooling flow cluster ($\dot M = 556~M_{\odot}$ yr$^{-1}$)
shows an initial rise in C76 and C745 before decreasing at larger radii,
indicating a drop in temperature. The C645 profile shows a slight, marginally
significant rise. At a temperature of 9.1 keV, abundance effects on the
colors are negligible, so a temperature drop is probably present in this
cluster.

{\it A2142}: This cluster was observed with both PSPC-B and PSPC-C, so the
data were analyzed separately. An initial rise in C76 and C745,
indicating a cooling flow ($\dot M = 350~M_{\odot}$ yr$^{-1}$),
levels out at large radii. The C645 profile is constant in the PSPB-B
observation, although the second C645 bin of the PSPC-C observation
is somewhat high. Henry \& Briel (1996)
found a constant temperature profile outside the cooling radius with the same
{\it ROSAT} data, although the temperature of their $2\farcm5-5^{\prime}$ bin
is unconstrained.

{\it A3112}: All three color profiles are constant within the errors,
despite the fact that this cluster harbors a strong cooling flow
($\dot M = 376~M_{\odot}$ yr$^{-1}$).
This cluster's temperature was found to decrease from 6 keV to 3 keV
with {\it ASCA} data (M98).

{\it A3571}: This cluster possesses a modest cooling flow
($\dot M = 72~M_{\odot}$ yr$^{-1}$). All three color profiles are constant with
small errors. M98 found this cluster's temperature to decrease by
a factor of 1.7 with {\it ASCA} data.

\subsection{$kT>5$ keV, $\dot M < 50~M_{\odot}$ yr$^{-1}$}
\label{ssec:highT-nocf}

{\it A119}: A marginally significant decrease of about 10\% in C76 and C745
is detected in this non-cooling flow cluster, indicating a temperature drop.
The constant C645 profile argues against a significant change in abundance
or absorption with radius. The innermost X-ray contours are somewhat irregular,
suggesting that this cluster is not completely relaxed and may have
undergone a recent merger (M98).

{\it A401}: C76 and C745 show a marginal increase with radius, while
C645 is constant. This is significant considering that M98 found the
temperature to decrease by a factor of 1.7 for this cluster, whereas
Fujita et al.\ (1996) found a constant temperature profile with the
same {\it ASCA} data. At 40\% metallicity, a drop in temperature from
10 keV to 6 keV (as was found with the {\it ASCA} data) would lead to
percentage drops in C76 and C745 of 4.6\% and 5.4\%, respectively.
However, this is inconsistent with the PSPC data at the 99.8\% and
99.98\% confidence levels for C76 and C745, respectively.
The cooling rate of this cluster is consistent with zero. Unfortunately,
A399 lies too far off-center from the A401 {\it ROSAT} observation to analyze
properly.

{\it A2256}: This cluster was the object of multiple pointings, but we choose
to analyze only the one long observation in which the center of the cluster
is in the center of the field of view. The other pointings were shorter and
offset, resulting in much of the emission falling outside the rib support
structure. In addition, the offset pointings were taken in the low gain mode
of the PSPC-B, whereas the long centered pointing was acquired with the
PSPC-C, preventing the combination of the data from the various pointings.
All three color profiles are constant within the errors,
indicating no temperature, metallicity, or absorption gradients in this
non-cooling flow cluster. Henry, Briel, \& Nulsen (1993) analyzed the
same {\it ROSAT} data (using just the PSPC-C data), and found an isothermal
profile out to 15$^{\prime}$. Markevitch (1996) found the temperature of
this cluster to decrease radially by a factor of two with {\it ASCA} data.

\begin{figure}[htbp]
\vskip6.30truein
\hskip0.3truein
\includegraphics{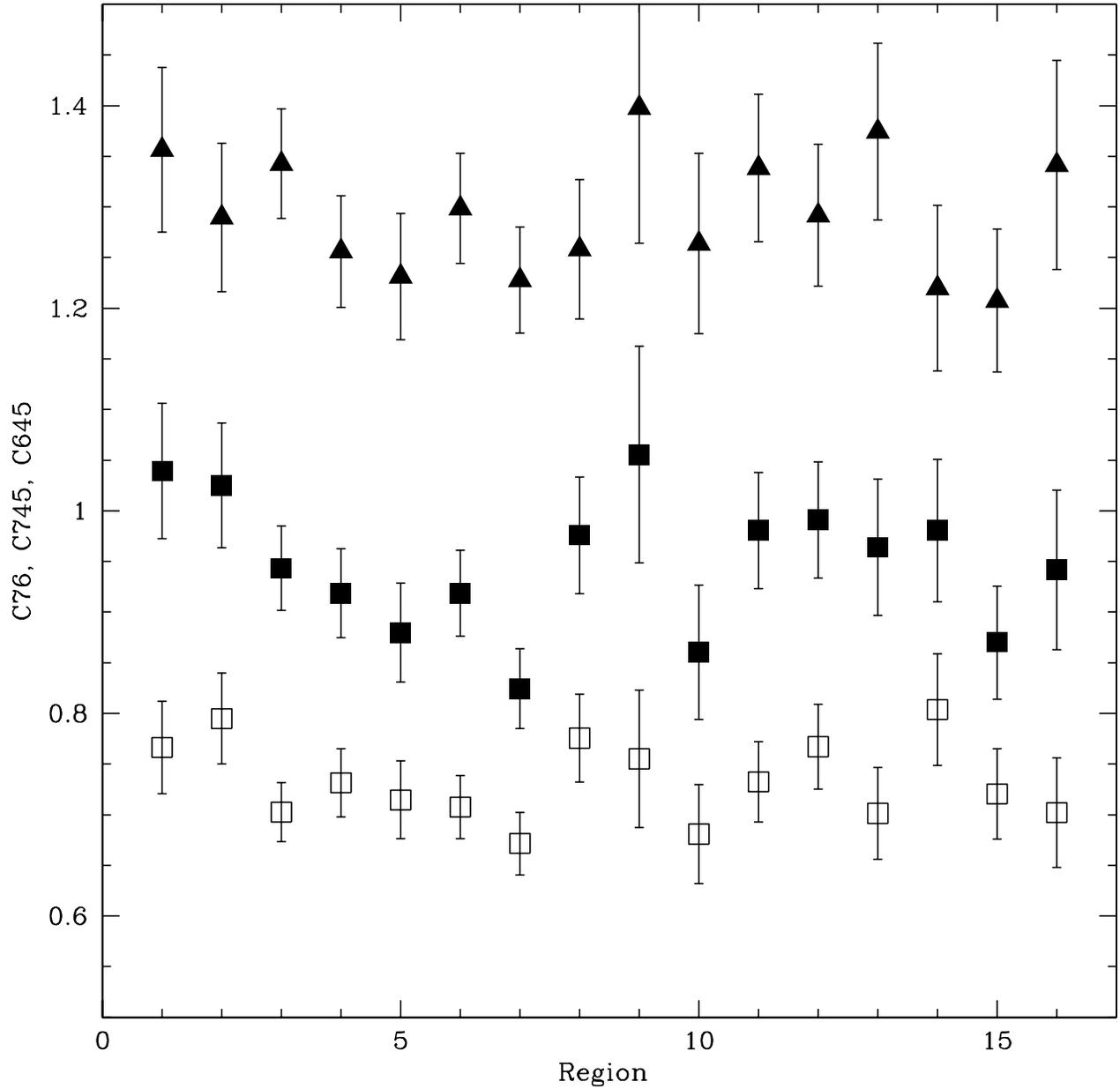}
\caption[A2256 Azimuthal Colors]{
Azimuthal color profiles for A2256. Regions 1--8 represent 45$^{\circ}$
sectors running counterclockwise from north for the inner $5^{\prime}$.
Regions 9--16 represent the same sectors for the $5^{\prime}-9^{\prime}$
radial bin.
\label{fig:a2256}}
\end{figure}

Briel \& Henry (1994) analyzed all five {\it ROSAT} PSPC pointings, and
found two hot spots ($kT>12$ keV) to the northeast and southwest of the center
from $5^{\prime}-9^{\prime}$. We determined the colors for the same regions
described by Briel \& Henry (1994), and the results are shown in
Figure~\ref{fig:a2256}. Regions 1--8 represent 45$^{\circ}$ sectors starting
north and running counterclockwise for the inner $5^{\prime}$. Regions 9--16
represent the same sectors for an annular bin of $5^{\prime}-9^{\prime}$.
Although the errors are large, there is a general agreement between the
two studies. For the inner bin, the cool spot of Region 7 and warm spot
of Region 8 are seen. For the outer bin, the hot spots of Region 9 and
Region 14 are seen. The hot spot of region 10 is missed, however, and in
fact appears as a cool spot in the color profile.

{\it A2319}: We have excluded data from a wedge between position angles
315$^{\circ}$ and 0$^{\circ}$ because of an asymmetry in the X-ray emission in
that region. Despite not having a cooling flow, the C76 and C745
profiles show a marginal initial increase. However, this trend is also seen in
C645, which is not sensitive to temperature. For all three colors,
the outer two bins are constant within the errors.

{\it A3158}: C76 and C745 show an initial decrease with radius before
leveling off, whereas C645 shows the opposite trend. It is not clear what
would cause this odd behavior of the color profiles other than a very steep
rise in temperature in the center of the cluster, since any radial trend in the
absorption should effect all three profiles in the same manner. This
cluster is not known to harbor an AGN.
The temperature profile appears constant outside of $2^{\prime}$.
The cooling rate of this cluster is consistent with zero.

{\it A3266}: C76 is flat in this non-cooling flow cluster, although the
C745 profile shows a marginally significant dip in the second spatial
bin. This dip is also present in C645 profile, indicating that it is
not a result of a change in temperature. A temperature drop from 10 keV
to 5 keV was found with {\it ASCA} data (M98).

{\it A3391}: The C76 and C745 profiles show a dip from the center to the
$2^{\prime}-6^{\prime}$ radial bin, while the C645 profile is constant.
Poor statistics preclude any definitive conclusion from being drawn on
this non-cooling flow cluster.

{\it Triangulum Australis}: This cluster also lies in a direction of high
Galactic hydrogen column density. Although the C76 profile is constant with
small error bars, C745 and C645 show a significant decrease at large
radii. The cooling rate of this cluster is consistent with zero.
The high (9.6 keV) temperature of this cluster precludes this trend
from being the result of an abundance gradient, and the constant C76 profile
argues against an absorption gradient. There appears to be no simple
explanation for the behavior of the C745 and C645 profiles at large radii,
other than some unknown source of excess band 4--5 emission. Since the
X-ray emission in this cluster extends to a large angular distance, we also
tried a larger background region. Extracting counts from an annular region of
$40^{\prime}-50^{\prime}$ for the background did not change the results.

\subsection{$kT<5$ keV, $\dot M > 50~M_{\odot}$ yr$^{-1}$}
\label{ssec:lowT-cf}

{\it A133}: This cool cluster shows a central drop in C76 and C745,
indicative of a cooling flow ($\dot M = 110~M_{\odot}$ yr$^{-1}$;
White, Jones, \& Forman 1997).
The outer two bins of C76 and C745 are consistent with a constant
temperature. The C645 profile shows an initial decrease, possibly
due to high metallicity in the center. This trend will be discussed
further in \S~\ref{sec:discussion}.

{\it A780 (Hydra A)}: C76 and C745 are constant outside of the innermost
radial bin, which contains a rather strong cooling flow
($\dot M = 264~M_{\odot}$ yr$^{-1}$). The C645 profile peaks in the center,
but levels out at larger radii.

{\it A2597}: All three color profiles are constant within the errors, although
the errors are quite large for the outermost bin. Sarazin \& McNamara (1997)
and Sarazin et al.\ (1995) found a large cooling flow
($\dot M = 344~M_{\odot}$ yr$^{-1}$) and excess absorption in this cluster with
the {\it ROSAT} PSPC and HRI, but detect a temperature drop and excess
absorption only in the inner $0\farcm75$. We see evidence for this in
the colors when we divide our innermost ($2^{\prime}$) bin into $0\farcm75$
segments. However, these features become diluted when the color
profiles are averaged over a $2^{\prime}$ bin.

{\it A4059}: C76 and C745 show an initial increase indicative of a cooling
flow ($\dot M = 130~M_{\odot}$ yr$^{-1}$), before
leveling off at larger radii. The C645 profile shows an
initial drop before leveling off, possibly due to
metallicity gradients that might not be evident in the C76 and
C745 profiles because of the cooling gas in this region. Huang \& Sarazin
(1998) found a constant temperature profile outside of the cooling radius
with the same {\it ROSAT} data.

{\it MKW3S}: Outside the cooling flow region
($\dot M = 175~M_{\odot}$ yr$^{-1}$), all three color profiles
are constant, albeit with rather large errors.

\subsection{$kT<5$ keV, $\dot M < 50~M_{\odot}$ yr$^{-1}$}
\label{ssec:lowT-nocf}

{\it A2589}: C76 and C745 show an initial increase indicative of a cooling
flow (although there is no published cooling rate for this cluster),
before leveling off at larger radii. The significant decrease in C645
coupled with a constant C76 and C745 profile would seem to rule out an
absorption gradient, but could be due to a metallicity gradient for this
low temperature ($\sim$4 keV; David, Jones, \& Forman 1996) cluster.
Both results are consistent with the findings of David et al.\ (1996)
who analyzed the same PSPC data and found a substantial metallicity
gradient in this cluster.

{\it A2657}: C76 is constant with small errors, while C745 and C645 show
an insignificant increase in this modest cooling flow cluster
($\dot M = 44~M_{\odot}$ yr$^{-1}$; White et al.\ 1997).

{\it A3532}: All three color profiles are constant within the errors
for this non-cooling flow cluster.

{\it A3562}: The C645 profile shows a significant decrease with radius,
indicative of an abundance or absorption gradient. The lack of a significant
decrease in C76 or C745 with radius would seem to rule out an
absorption gradient. In fact, an increasing C76 profile and a slightly
decreasing C745 profile indicates a large decrease in metallicity with
increasing radius (see Figure~\ref{fig:model_temps}). The effect of
metallicity on the colors is measurable in this case because of the cluster's
low temperature ($kT=4.2$ keV; Henry \& Briel 1996). The nearly constant C745
profile indicates a constant temperature profile, apparently unaffected by
the cluster's small cooling flow ($\dot M = 37~M_{\odot}$ yr$^{-1}$).

\section{Analysis and Discussion of Composite Color Profiles}
\label{sec:discussion}
\subsection{Temperature Gradients} \label{ssec:temp_grad}

Individually, the lack of a simultaneous decrease in the C76 and C745
profiles (the colors sensitive to changes in temperature) suggests that the
radial temperature profile is nearly constant for most clusters, at least
outside the cooling region. However, in many cases the
statistics are not good enough to definitively rule out a temperature
gradient. In order to reduce the statistical uncertainty, we have
calculated color profiles for our sample as a whole. For each cluster,
counts were extracted from three concentric annular bins. The size of the
bins were chosen to be the same fraction of the virial radius for
each cluster, in order to compare physically similar regions of each
cluster regardless of size or distance. The outer radii of each bin were
chosen to be 7.9\%, 15.8\%, and 35.5\% of the virial radius. This choice of
bin size was selected so that the extraction region extended to at least
$1\farcm5$ (to contain all or nearly all cooling flow emission; only two
clusters had innermost annuli of less than 2$^{\prime}$ in extent) and the
outermost bin to no larger than 18$^{\prime}$ (to stay within the rib support
structure) for all clusters. Note that this
extraction method was not used to derive the individual profiles shown
in Figure~\ref{fig:highT-cf}--\ref{fig:lowT-nocf}; for the individual profiles 
we attempted to match the extraction regions with those chosen by the 
authors of the respective ASCA clusters for a direct comparison.  We assume the typical relation between
cluster virial size (the radial scale encompassing a mean density
contrast of 180 relative to the critical density) and temperature 
with normalization $r_{virial} = 3.9 (T/10 {\rm keV})^{1/2}$ Mpc, 
assuming a Hubble constant of 50 km s$^{-1}$ Mpc$^{-1}$.

For each cluster, the color profiles were normalized by the global value
for that color (i.e., C76$(r)$/$<$C76$>$, C745$(r)$/$<$C745$>$,
C645$(r)$/$<$C645$>$), so that the color profiles for each cluster varied
around a value of one. The color profiles were averaged together,
with each cluster given equal weights. The clusters were also averaged by
weighting each cluster according to the number of counts that cluster
contributed to the sum total of counts (calculated separately for each spatial
bin). The two procedures yielded results that agreed to within 1\% for each
spatial and color bin, and was less than 0.5\% for a majority of bins.
Since the equal weighting average yielded somewhat higher errors (given
by the uncertainty in the mean for the cluster sample), we adopt the equal
weighting scheme to be conservative.

\begin{figure}[htbp]
\vskip6.30truein
\hskip0.3truein
\includegraphics{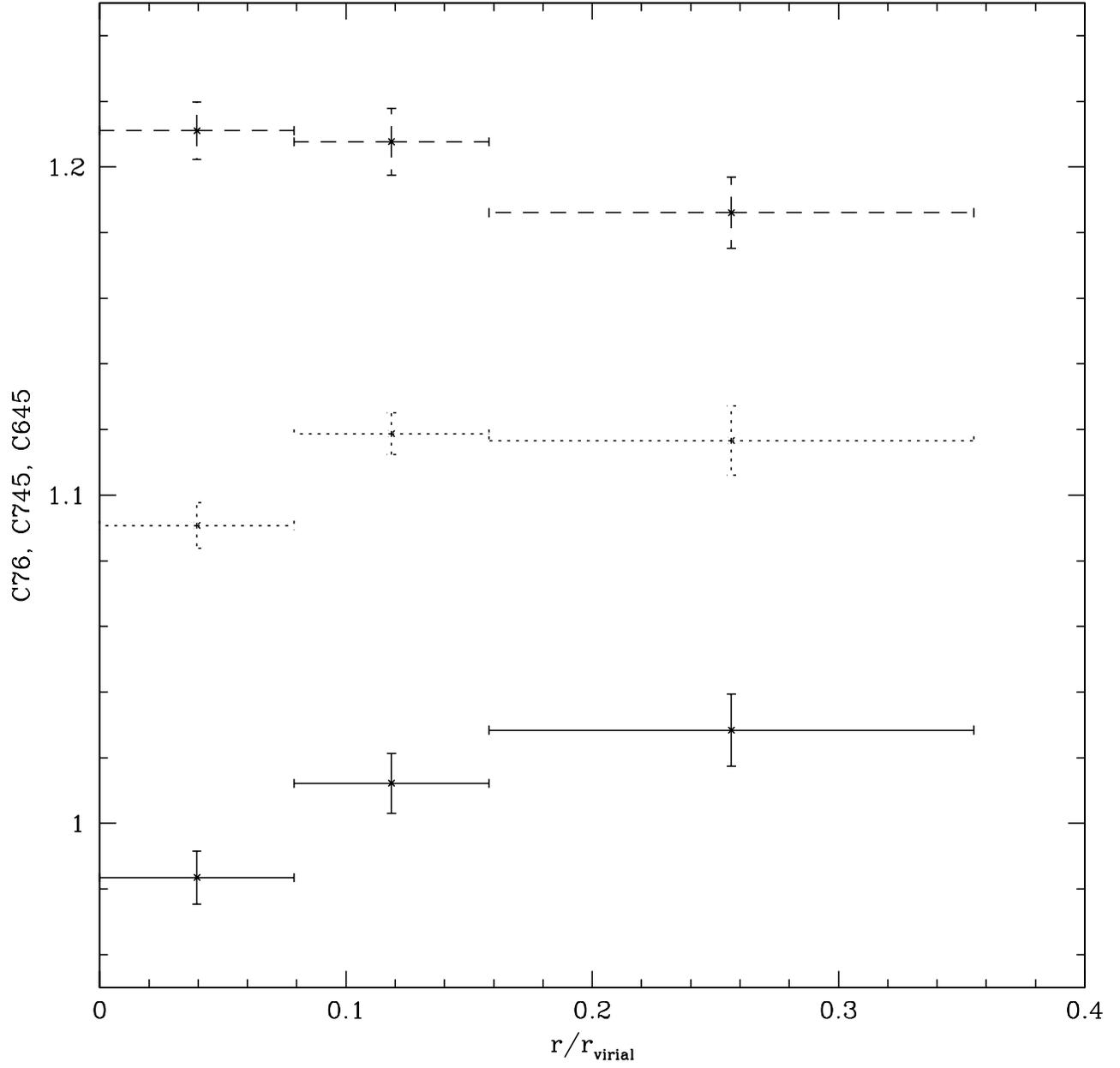}
\caption[Composite Color Profile - All Clusters]{
Composite radial color profiles for all 26 clusters in the sample. The
C76 profile is shown in solid lines, C745 is shown in dotted lines, and
C645 is shown in dashed lines. The errors represent the deviation in the
mean for the sample.
\label{fig:composite_all}}
\end{figure}

The composite color profiles for all clusters in the sample is shown
in Figure~\ref{fig:composite_all}. For display
purposes, we have multiplied the C745 profile (dotted line) by a factor of
1.1 and the C645 profile (dashed line) by 1.2. The outer two bins for both
C76 and C745 show remarkable agreement, and the errors are 1\% or less (given
by the uncertainty of the mean of the sample).
Since these two colors are sensitive to changes in
temperature, this strongly suggests isothermality in the outer regions
of clusters, at least out to 35\% of the virial radius. At 10 keV (5 keV), a
drop in temperature from bin 2 to bin 3 of 20\% (12\%) is ruled out at
the 99\% confidence level. The innermost bin
of these two profiles is noticeably lower than the other bins, suggesting a
temperature drop in the center. This is to be expected considering that
the innermost bin is similar in size to the cooling radius of clusters
that possess cooling flows. This drop is not seen in the inner bin of the
composite C645 profile, since C645 is insensitive to changes in temperature.
However, this color is sensitive to absorption and metallicity gradients,
both of which are expected (if present) to lead to a softening of the
profile with increasing radius for physically realistic situations.

\begin{figure}[htbp]
\vskip6.30truein
\hskip0.3truein
\includegraphics{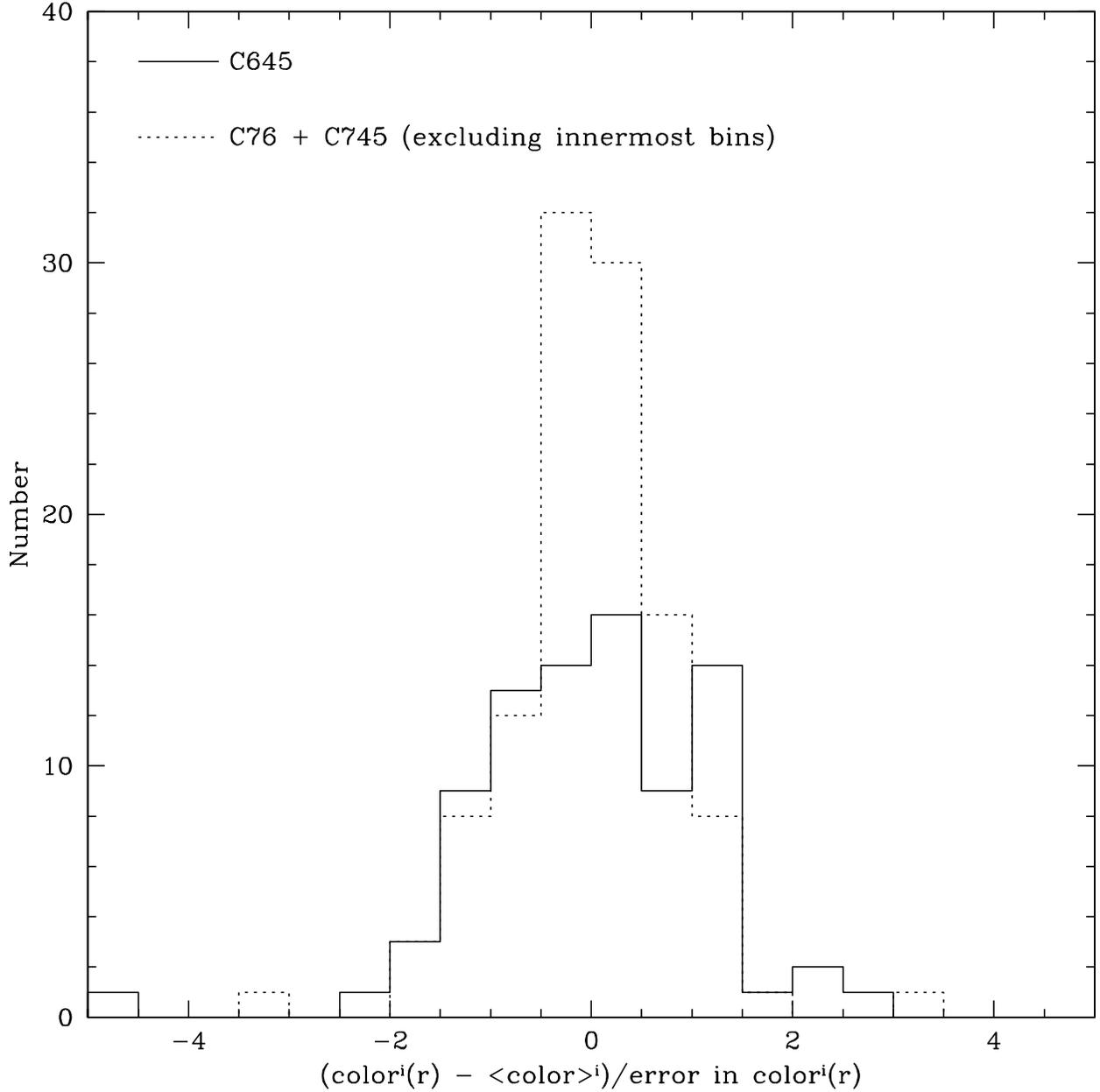}
\caption[Histogram of C645 and C76+C745 Colors]{
Distribution of (color$^i$($r) - <$color$>^i$)/(error in color$^i$($r$))
values for C645 (solid) and C76 + C745 (dotted) for each spatial bin of
each individual cluster $i$ in the sample.
For C76 + C745, the innermost spatial bins were
excluded since they contain cooling gas in many of the clusters. For both
distributions, the shape is approximately Gaussian with a mean near zero
and a dispersion close to unity. For C645 this implies that our contamination
source removal and background subtraction are accurate. For C76 + C745,
this result implies that the consistency between the outer two spatial bins
is not the result of merging together rapidly increasing and rapidly
decreasing individual color profiles.
\label{fig:histogram}}
\end{figure}

The main purpose for calculating the temperature-insensitive C645 profiles was
to have an indicator as to whether some unknown systematic effect was biasing
our results by introducing a spurious gradient to the color profiles.
As shown above, this color profile should be flat (neglecting metallicity
effects), so any radially-dependent systematic effect that might be
biasing the temperature-sensitive C76 and C745 colors should be easily
detectable in C645. To test this, we have calculated the significance of the
deviations of the C645($r$) values from the global C645 value for each
cluster for each spatial bin in the sample, i.e.,
(C645$^i(r) - <$C645$>^i$)/(error in C645$^i$($r$)), for each cluster
$i$. If there is no
unknown systematic effect, one would expect a Gaussian distribution
with zero mean and unit dispersion. Figure~\ref{fig:histogram} shows
that the resulting histogram is approximately Gaussian in shape. The
mean and standard deviation of the distribution are $-0.009 \pm 0.12$
and $1.13 \pm 0.09$, respectively. Thus, the distribution is only slightly
wider than expected, indicating that there is no systematic effect biasing
the radial color profiles.
\begin{figure}[htbp]
\vskip6.30truein
\hskip0.3truein
\includegraphics{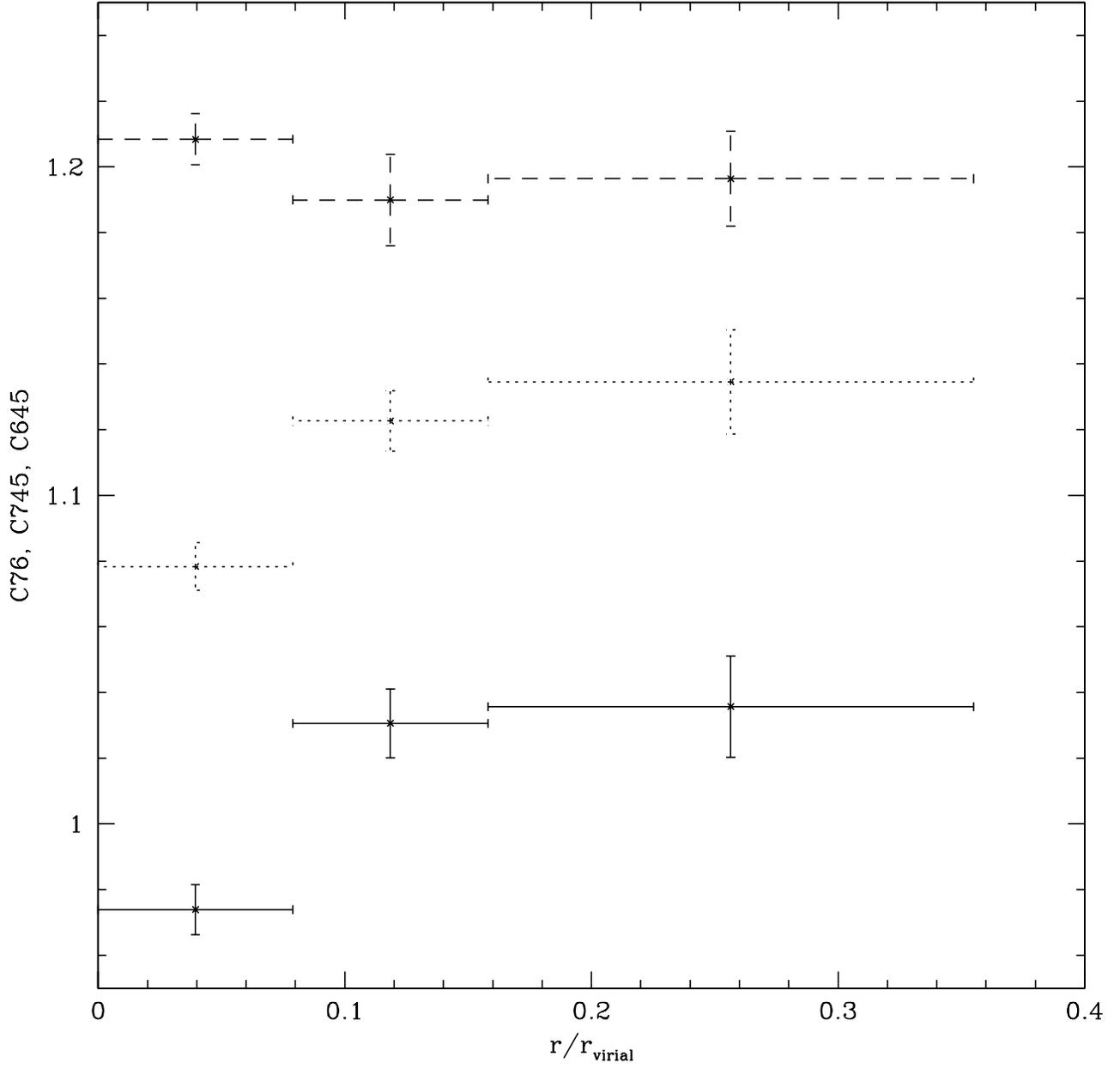}
\caption[Composite Color Profile - Cooling Flow Clusters]{
Composite radial color profiles for all cooling flow galaxies with
cooling rates higher than $50~M_{\odot}$ yr$^{-1}$. The notation is the same
as that in Figure~\protect\ref{fig:composite_all}.
\label{fig:composite_cf}}
\end{figure}
\begin{figure}[htbp]
\vskip6.30truein
\hskip0.3truein
\includegraphics{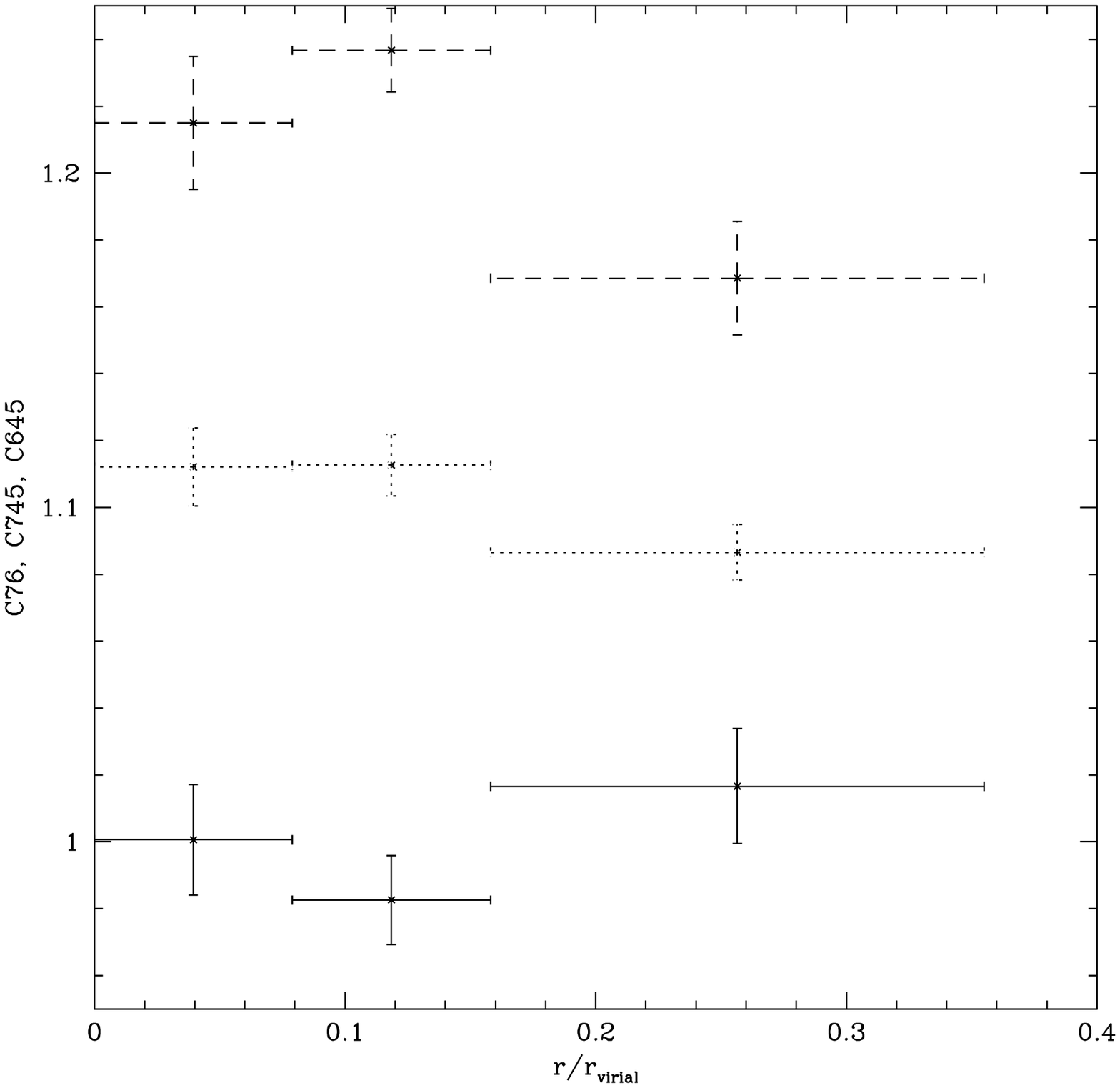}
\caption[Composite Color Profile - Non-Cooling Flow Clusters]{
Composite radial color profiles for all non-cooling flow galaxies.
The notation is the same as that in Figure~\protect\ref{fig:composite_all}.
\label{fig:composite_nocf}}
\end{figure}

The agreement in the outer two bins of the composite C76 and C745 profiles
is not the result of averaging together rapidly increasing profiles
with rapidly decreasing individual color profiles.
The same statistical test used to determine the robustness of
the constancy of C645 was used on the outer two spatial bins of C76 and
C745 for each cluster (we have omitted the inner cooling flow
spatial bin, and determined the
global color using only the outer two spatial bins). The results are shown
in Figure~\ref{fig:histogram}. Again, the histogram is Gaussian, with a
mean of $-0.014 \pm  0.077$ and a dispersion of $0.82 \pm 0.06$. Obviously, if
we were merging together strongly increasing and strongly decreasing
individual color profiles, the width of the Gaussian would be much more than
unity.

Next, we separated the sample into two groups according to whether the cluster
possessed a substantial cooling flow ($>50~M_{\odot}$ yr$^{-1}$) or not
to verify that the drop in C76 and C745 in the innermost bin was due
to cooling gas at lower temperatures. The color profiles for the cooling
flow subsample are shown in Figure~\ref{fig:composite_cf}. The drop in C76
and C745 is very pronounced, yet the excellent agreement of the outer
two radial bins remains. The C645 profile is constant within the errors.
The sharp drop in the innermost bin of C76 and C745 disappears in the
non-cooling flow subsample (Figure~\ref{fig:composite_nocf}). Note that
A2589 is not included in this subsample, since no cooling rate was
available from the literature. Here, C76 is constant within the errors and
C745 shows only a very small drop in the outermost bin, again indicating a
lack of a temperature gradient. There is a curious 5\% drop in the outermost
bin of the C645 profile. This feature disappears, however, upon the removal
of three clusters -- A3532, A3562, and Triangulum Australis. Irregularities
in the C645 profiles of the latter two clusters were discussed in
\S~\ref{sec:color_profiles}.

One complication that could lead to differences in the measured temperatures
between {\it ROSAT} and {\it ASCA} is the presence of multi-temperature
gas. In that event, {\it ROSAT} might measure a lower temperature
than {\it ASCA}, because {\it ROSAT} would be more sensitive to the low
temperature component of the multi-temperature gas than {\it ASCA}, which
would be sensitive to higher temperature gas. However, such multi-temperature
gas might only be expected to occur within the cooling flow region of
clusters (the first spatial bin), and should not occur past the
cooling radius. This would have no effect on the outer two spatial bins,
where the case for isothermality is strongest.

Given the small uncertainties of the composite color profiles,
changes in the temperature of a factor of two should have been detected,
since changes of this magnitude would lead to changes in the C76 and
C745 profiles of at least 7\%. It therefore appears that outside the
cooling radius, the gas in clusters of galaxies is largely isothermal.

\subsection{Abundance Gradients} \label{ssec:abun_grad}

Inspection of Figures~\ref{fig:lowT-cf} and \ref{fig:lowT-nocf} reveals
that many of the low temperature clusters show a higher central value of
C645 than the rest of the cluster, most notably A133, A780, A2589, A3562,
and A4059. Since this color is not sensitive to temperature, it cannot be
due to a change in temperature. Nor can it be due to an absorption gradient,
as this would also be evident in C76 and C745 as well.
However, a high central metallicity can be responsible for this,
since C645 is much more sensitive to metallicity than C76 and C745,
especially at low temperatures (see Figure~\ref{fig:model_temps}).
\begin{figure}[htbp]
\vskip6.30truein
\hskip0.3truein
\includegraphics{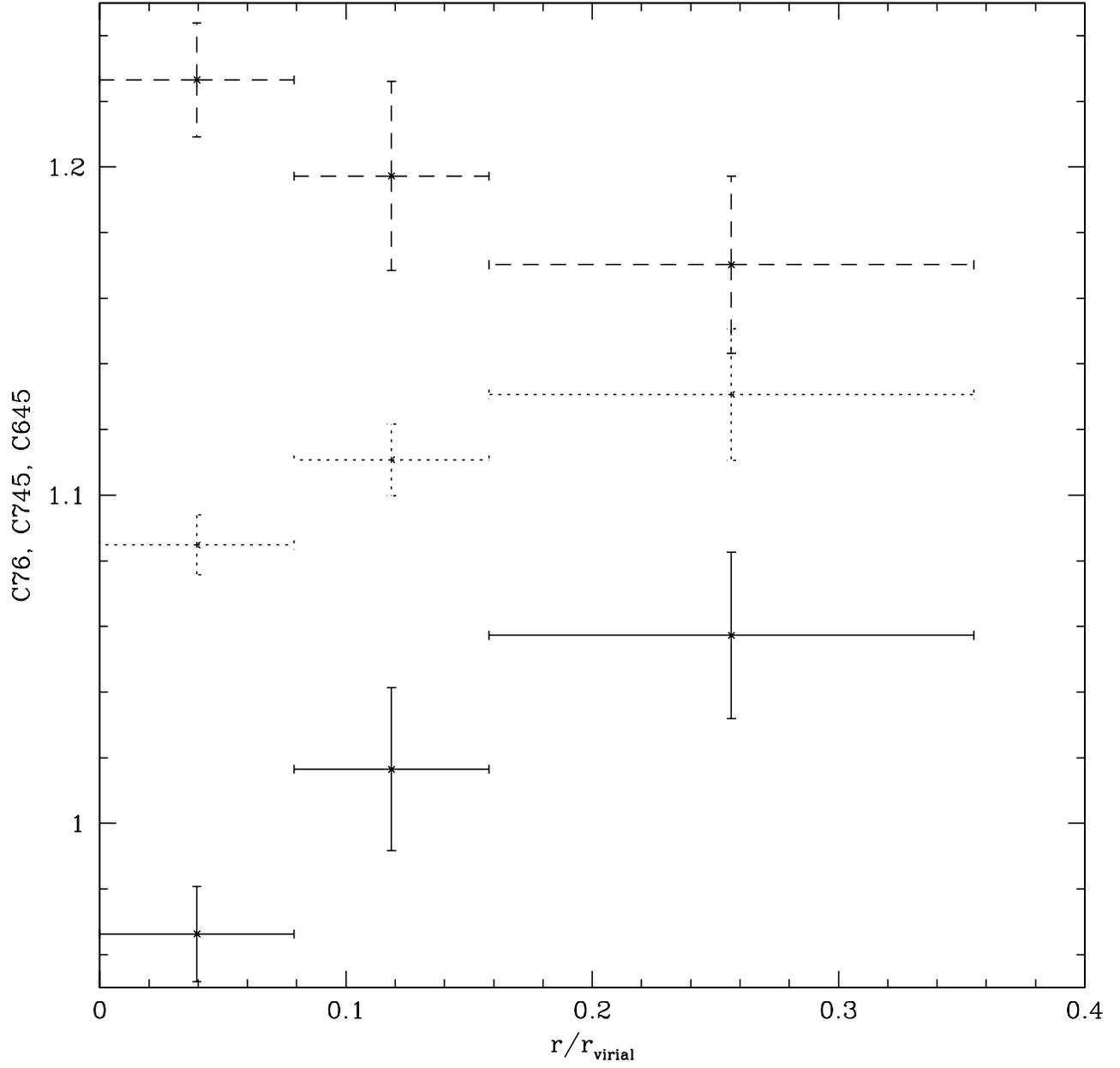}
\caption[Composite Color Profile - Cool Clusters]{
Composite radial color profiles for all clusters with $kT<5$ keV.
The notation is the same as that in Figure~\protect\ref{fig:composite_all}.
A definite decline in C645 is evident, indicative of a negative radial
abundance gradient.
\label{fig:composite_lowT}}
\end{figure}
\begin{figure}[htbp]
\vskip6.30truein
\hskip0.3truein
\includegraphics{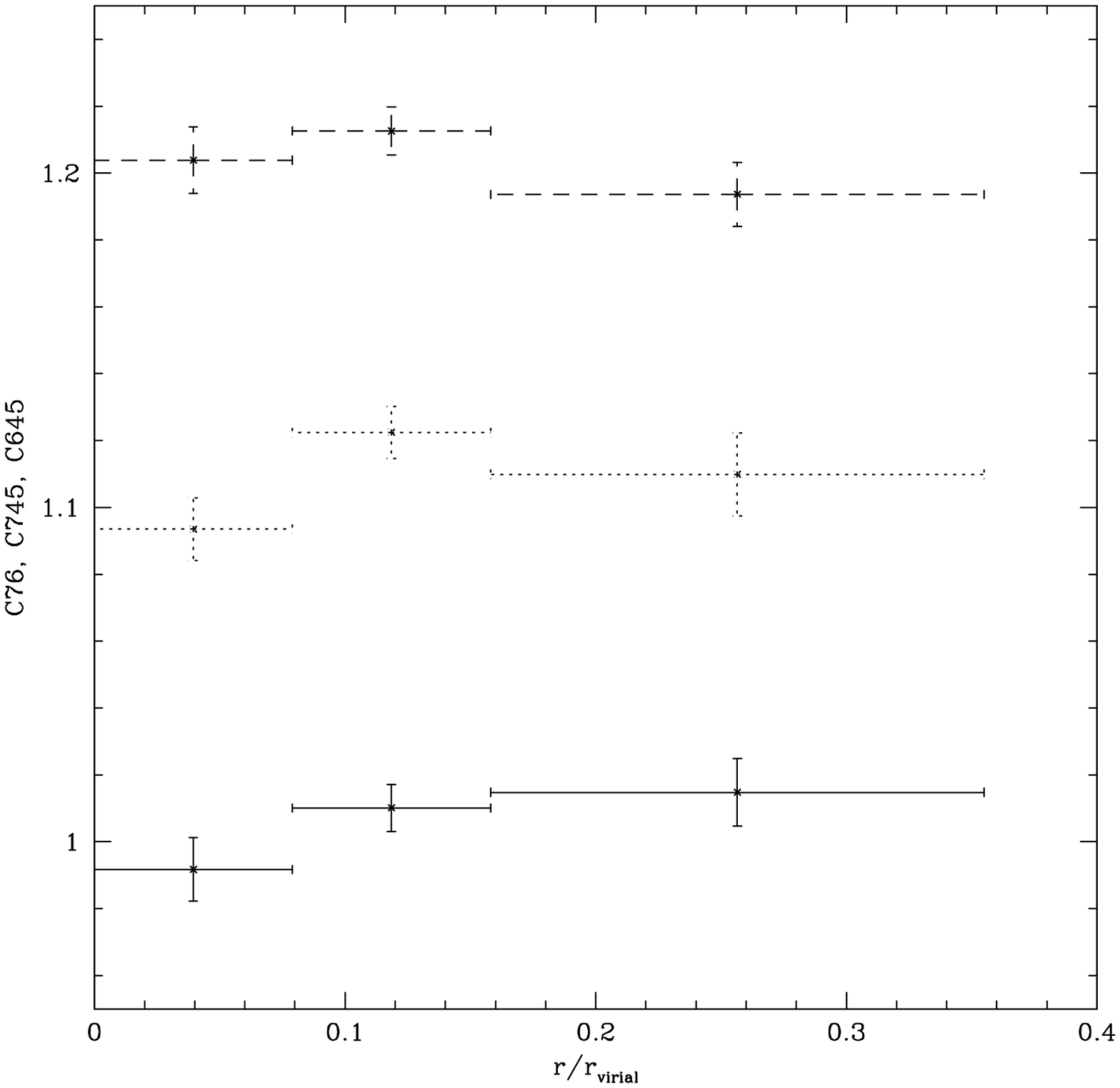}
\caption[Composite Color Profile - Hot Clusters]{
Composite radial color profiles for all clusters with $kT>5$ keV.
The notation is the same as that in Figure~\protect\ref{fig:composite_all}.
The decline in C645 seen in the low temperature clusters is not seen here,
since C645 is not sensitive to abundance changes at higher temperatures.
\label{fig:composite_highT}}
\end{figure}

We have combined the color profiles of
all clusters with $kT < 5$ keV in Figure~\ref{fig:composite_lowT}. The C76 and
C745 profiles rise in the central regions since there are cooling flow clusters
in this subsample. The composite C645 profile shows a noticeable decreasing
trend with radius, indicative of a possible negative radial abundance
gradient. This trend remains upon the removal of A3562 and A3532, indicating
that the trend is not the result of a couple of nonrepresentative clusters.
The decrease in C645 from the innermost bin to the outermost bin is
4.6\%. At 4 keV, a decrease of this magnitude could be caused by a decrease
in metallicity from 60\% solar to 20\% solar, or equivalently from 40\%
solar to 6\% solar. Given that there is also
a temperature decrease in the center, the decrease in metallicity may be
even greater to account for the observed decrease in C645.
As a comparison, we show the composite profiles for clusters with
$kT > 5$ keV in Figure~\ref{fig:composite_highT}. Here, the C645 profile is
nearly flat, and in fact becomes flat if Triangulum Australis is removed from
the subsample. Abundance gradients may be present in the
high temperature clusters just like their low temperature counterparts,
but as Figure~\ref{fig:model_temps} shows C645 is insensitive to
metallicity above 5 keV. It should also be noticed that the central
decrease in C76 and C745 is less pronounced in the high temperature
subsample than the low temperature subsample, since these colors are less
sensitive to changes in temperatures brought on by the presence of cooling
flows at high temperatures.

Previous studies have yielded conflicting results concerning the presence
of metallicity gradients in clusters. Whereas some clusters show no
evidence for the presence of a metallicity gradient such as Ophiuchus
(Matsuzawa et al.\ 1996), A1060 (Tamura et al.\ 1996), A399, and A401
(Fujita 1996), other clusters such as AWM7 (Ezawa et al.\ 1997), A496
(Hatsukade et al.\ 1996), and Centaurus (Ikebe 1995) exhibit significant
metallicity gradients. Allen \& Fabian (1998)
found that the emission-weighted metallicities of cooling flow clusters
were on average 1.8 times higher than those of non-cooling flow clusters,
suggesting that metallicity gradients that were present in cooling flow
clusters but absent in non-cooling flow clusters was responsible for the
discrepancy. The strongest example in our sample of a cluster having a
metallicity gradient is A3562, which has a low ($37~M_{\odot}$ yr$^{-1}$)
cooling rate. From our sample it would appear that some, but not all
clusters exhibit a central peak in metallicity.

\section{Comparison to Cosmological Models} \label{sec:models}

Next, we compare the observed colors to those predicted from a series
of theoretical temperature profiles derived from hydrodynamic cluster
simulations for various cosmological models. We have used
the temperature profiles from six cosmological models presented in
Evrard, Metzler, \& Navarro (1996).
The models are for cosmologies with $\Omega=1$ with energy feedback
and mass ejecta from galaxies within the cluster (EJ), and without feedback
(2F) (Metzler 1995), $\Omega=1$ using an independent code from
Navarro, Frenk, \& White (1995) (NFW), and three models from
Evrard et al.\ (1993), with a standard CDM model with $\Omega=1$ (EdS),
an open universe with $\Omega=0.2$ (Op2), and a low-density universe
with a flat cosmology with $\Omega=0.2$ and a cosmological constant
$\lambda=0.8$ (Fl2). Further details concerning the models can be found
in Evrard et al.\ (1996), and references therein.
\begin{figure}[htbp]
\vskip6.30truein
\hskip0.3truein
\includegraphics{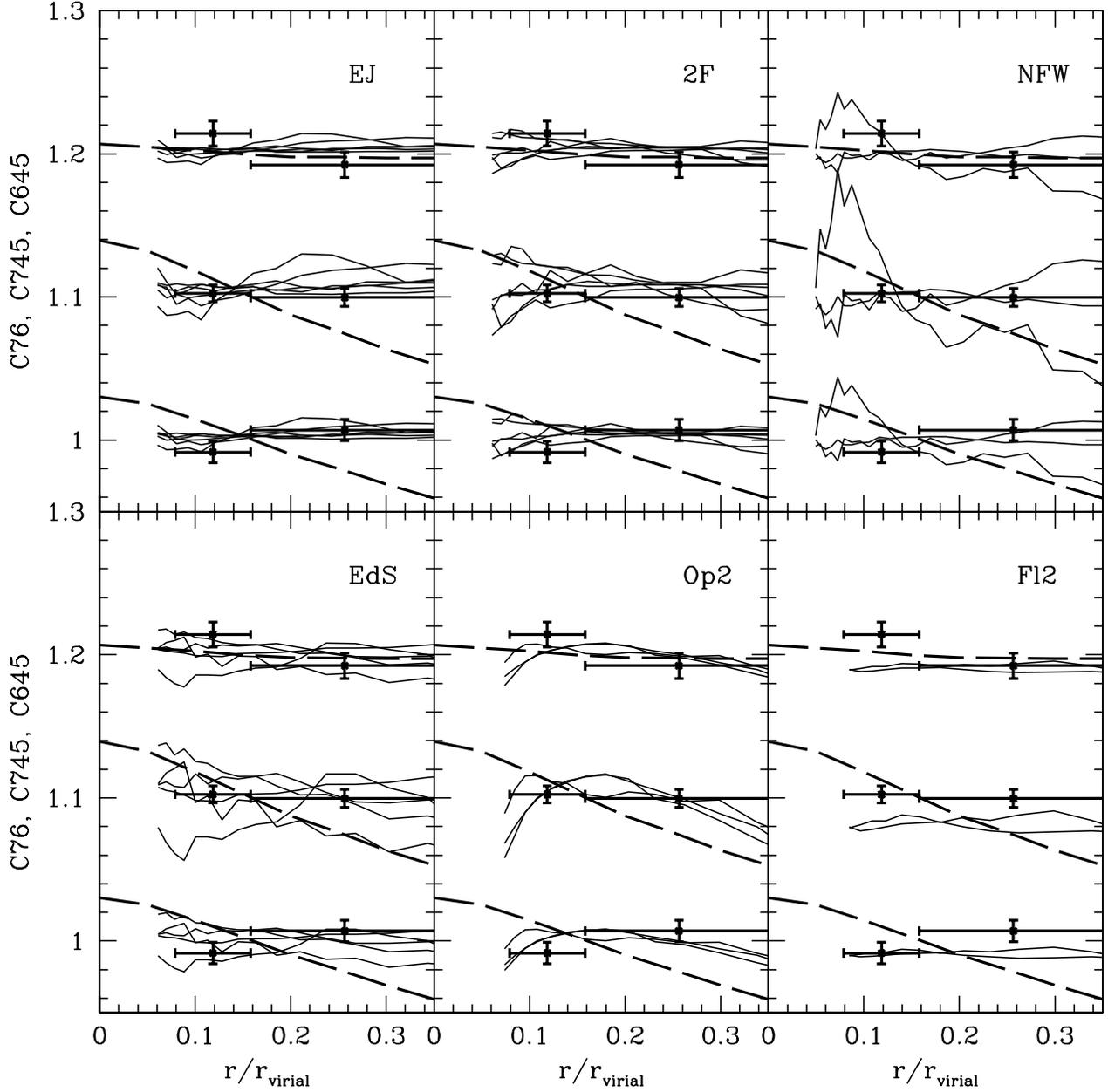}
\caption[Colors Derived From Simulations]{
Comparison of the colors of the temperature profiles from six
cosmological models taken from Evrard et al.\ (1996; solid lines) with
those of the outer two bins of the composite profile for all clusters
in our sample and with the best-fit polytropic relation between temperature
and radius from M98 ($\gamma=1.24$; long dashed line). For clarity, the
C745 and C645 profiles have been multiplied by 1.1 and 1.2, respectively.
For each cosmology, the models cover a range of temperatures and masses.
The polytropic relation of M98 is inconsistent with both the models and
the {\it ROSAT} data.
\label{fig:gus_models}}
\end{figure}

The models cover a range of temperatures and masses, and the colors of
these models are presented in Figure~\ref{fig:gus_models}. We
have excluded all runs where the global temperature was less than 3.5 keV,
since there are no such low temperature clusters in our sample of observed
clusters. The colors were calculated for each spatial bin, and normalized
to the colors for the global temperature for that run. Note that we have
converted the $r_{500}$ notation used in Evrard et al. (1996) to $r_{virial}$.
Once again the normalized C745 and C645 profiles were multiplied by 1.1 and 1.2,
respectively, for clarity. The model colors have been plotted out to
35\% of the virial radius to match the data. Within each panel,
the colors from the normalized composite profile are shown for comparison.
This time the innermost bin (which contains cooling gas) has been omitted
from the normalization procedure. It is evident that the agreement of
the colors in the outer two spatial bins of C76 and C745 is predicted
by the models. The one exception is the Fl2 model, but this is possibly due to
a misnormalization of the temperature profile in Evrard et al.\ (1996); the
Fl2 panel of Figure~5 of that paper shows that the temperature profile
$T(r)/T_X$ peaks at a value of one, rather than fluctuating around a value
of one as a normalized profile should. Still, the color profile is flat,
even if the normalized color value is less than one.
Unfortunately, the similarity of the models at radii
less than 35\% of the virial radius precludes the use of the data
from distinguishing among the different cosmologies presented here.

The agreement between the models and the {\it ROSAT} data disagrees
with the work of M98 who used {\it ASCA} data to derive normalized
temperature profiles of a sample of clusters. M98 found that, on average,
the temperature decreased from about 1.2 times the global temperature
at 5\% of the virial radius to 0.7 times the global temperature at 35\%
of the virial radius. M98 fit the temperature profiles with a polytropic
equation of state, which leads to
\begin{equation}
T(r) \propto \left( 1 + \frac{r^2}{a_x^2} \right)^{-3\beta (\gamma-1)/2} ,
\end{equation}
where $r$ is the projected distance, $a_x$ is the core radius, $\beta$
has its usual meaning in the context of isothermal $\beta$-models, and
$\gamma$ is the polytropic index. Using values of $a_x=0.092r_{virial}$
and $\beta=0.67$ (typical values for a 7 keV cluster; Jones \& Forman 1997),
M98 found a best
fit polytropic index of $\gamma=1.24^{+0.20}_{-0.12}$ (90\% confidence levels).
We have calculated the colors for a 7 keV cluster that follows this
radial profile with $\gamma=1.24$ and plotted them in 
Figure~\ref{fig:gus_models}
(long dashed line). The resulting color profiles are incongruous with
both the {\it ROSAT} data and the models.

M98 claim that two cosmological models predict a significant temperature
decrease that matches their {\it ASCA} data: that of Bryan \& Norman (1997)
which employs a Eulerian simulation of an $\Omega=1$ cluster without
cooling, and that of Katz and White (1993), which employs a Lagrangian
simulation of an $\Omega=1$ cluster with cooling. The Bryan \& Norman (1997)
simulations show a steep drop in temperature in the inner regions, but were
performed only on a single cluster. Other simulations by the
same authors (Bryan \& Norman 1998) performed on five clusters find
flatter temperature profiles. For their cold dark matter model, the
temperature profile falls very slowly out to 0.7$r_{virial}$. For their
cold plus hot dark matter model, the temperature profile decreases
$\sim$15\% from 15\% to 35\% of the virial radius for their steepest
profile, and only a few percent for their shallowest profile. All the hot plus
dark matter profiles are consistent with the cold dark matter model profiles
at radii $>35\%$ of the virial radius, where the cold dark matter models begin.
The drop in temperature out to 0.7$r_{virial}$ is
significantly less steep than in Bryan \& Norman (1997).
The steep temperature gradient
seen by Katz \& White (1993) may have been caused by spurious ram pressure
heating of the hot ICM by plunging galaxies in their model, and may not
represent a physically plausible situation.

The reliability of such simulations has been empirically tested by
Frenk et al.\ (1998).  They compare the structural properties of a
single cluster in a cold dark matter $\Omega=1$ universe evolved by twelve
different cosmological codes. In general, models using a Smoothed
Particle Hydrodynamics (SPH) technique (used to generate all but the ``NFW''
panel in Figure~\ref{fig:composite_all}) show a flat or slowly declining
temperature profile within 35\% of the virial radius (consistent with
the {\it ROSAT} data), while models using a mesh grid method exhibit a
significantly decreasing radial trend in the temperature.
The steepest profile is from the Bryan \& Norman simulation, which
was taken from Bryan \& Norman (1997)
and not from the simulations of Bryan \& Norman (1998). It shows a decrease
of $\sim$60\% in temperature from 0.05 to 0.35$r_{virial}$. The profile
from the simulations of Cen show a similar drop in temperature. However,
these are the only two simulations that extend to radii less than
0.2$r_{virial}$ that show such a significant temperature decrease.
It should also be noted that these are 3-D temperature profiles and would
appear less steep in projection.

It is debatable how wide a conclusion can be drawn
from comparison of a single model realization, but the slight differences
between the two codes in the very central regions may be real and 
perhaps reflect a generic condition between solutions obtained with
Eulerian (grid) versus Lagrangian (SPH) methods.  
Not all Lagrangian SPH studies show completely flat central temperature
profiles though.  With a set of low density universe simulations, 
Eke, Navarro \& Frenk (1998) show that the mean temperature profile 
derived from the ten most massive objects exhibits a $20\%$ drop
between $0.1$ and $0.35 r_{virial}$. Still, this is a significantly
shallower drop than is measured by M98.

Making interpretation more complicated is the fact that the current
generation of simulations is not likely to be modeling the physics of the
inner cluster correctly at the $\sim$5--10\% level.
Possible complications include a multi--phase central structure, 
magnetic fields, metallicity gradients (which affect cooling rates),
galactic ram pressure and small scale turbulence. However, these
potential complications are of less significance outside the cooling
radius, where the case for isothermality is strongest.

In conclusion, although a few simulations exist that predict a steep
temperature drop in the inner regions of clusters, most published models
that have sufficient resolution to probe the inner 10\% of the virial
radius predict a nearly flat temperature profile, in agreement with the
{\it ROSAT} data, but in conflict with M98.

\section{Final Comments and Future Work} \label{sec:comments}

Both individually and cumulatively, our color profiles suggest that 
the temperature profiles of the hot gas within clusters of galaxies
are constant with radius outside of the cooling radius. More conclusive
evidence supporting this claim will be provided by {\it AXAF}, with its
wide energy bandpass and excellent spatial resolution.  Given the
large discrepancy in the measured temperature profiles of A401 and A399
with {\it ASCA} using different methods by previous authors,
these two clusters would be an ideal test case for {\it AXAF}. {\it AXAF}
observations of clusters should also resolve the issue of abundance
gradients in clusters, as has been indicated in this and other studies.

\acknowledgments

We thank Joe Mohr, Pat Henry, and Ulrich Briel for many very useful
comments and discussions.
This research has made use of data obtained through the High Energy
Astrophysics Science Archive Research Center Online Service,
provided by the NASA/Goddard Space Flight Center.
This work has been supported by NASA grants NAG5-3247, NAG5-7108, and
NAG5-3401.

\end{document}